\documentclass[english,10ppt,nofootinbib]{revtex4}
\usepackage[T1]{fontenc}
\usepackage[latin9]{inputenc}
\usepackage{color}
\usepackage{float}
\usepackage{amsmath}
\usepackage{amssymb}
\usepackage{graphicx}

\makeatletter
\@ifundefined{textcolor}{}
{%
 \definecolor{BLACK}{gray}{0}
 \definecolor{WHITE}{gray}{1}
 \definecolor{RED}{rgb}{1,0,0}
 \definecolor{GREEN}{rgb}{0,1,0}
 \definecolor{BLUE}{rgb}{0,0,1}
 \definecolor{CYAN}{cmyk}{1,0,0,0}
 \definecolor{MAGENTA}{cmyk}{0,1,0,0}
 \definecolor{YELLOW}{cmyk}{0,0,1,0}
}

\usepackage{babel}

\usepackage{babel}

\makeatother

\usepackage{babel}
\begin{document}

\title{$f\left(R,\nabla_{\mu_{1}}R,\dots,\nabla_{\mu_{1}}\dots\nabla_{\mu_{n}}R%
\right)$ theories of gravity in Einstein frame: A higher order modified Starobinsky  inflation model in the Palatini approach}
\author{R. R. Cuzinatto$^{1}$}
\email{rodrigo.cuzinatto@unifal-mg.edu.br}
\author{C. A. M. de Melo$^{1,2}$}
\email{cassius.melo@unifal-mg.edu.br}
\author{L. G. Medeiros$^{2,3}$}
\email{leogmedeiros@ect.ufrn.br}
\author{P. J. Pompeia$^{4}$}
\email{pompeia@ita.br}
\affiliation{$^{1}$Instituto de Ci\^encia e Tecnologia, Universidade Federal de Alfenas,
Rodovia Jos\'e Aur\'elio Vilela, 11999, Cidade Universit\'aria, CEP 37715-400 Po\c cos de Caldas, Minas Gerais, Brazil}
\affiliation{$^{2}$Instituto de F\'isica Te\'orica, Universidade Estadual Paulista, Rua Bento
Teobaldo Ferraz, 271, Bloco II, P.O. Box 70532-2, CEP 01156-970 S\~ao Paulo, S\~ao Paulo, Brazil}
\affiliation{$^{3}$Escola de Ci\^encia e Tecnologia, Universidade Federal do Rio Grande do
Norte, Campus Universit\'ario, s/n \textendash{} Lagoa Nova, CEP 59078-970
Natal, Rio Grande do Norte, Brazil}
\affiliation{$^{4}$Departamento de F\'isica, Instituto Tecnol\'ogico da Aeron\'autica, Pra\c ca
Mal. Eduardo Gomes, 50, CEP 12228-900 S\~ao Jos\'e dos Campos, S\~ao Paulo, Brazil}

\begin{abstract}
In Cuzinatto \textit{et al.} [Phys. Rev. D 93, 124034 (2016)], it has been
demonstrated that theories of gravity in which the Lagrangian includes terms
depending on the scalar curvature $R$ and its derivatives up to order $n$ ,
i.e. $f\left( R,\nabla _{\mu }R,\nabla _{\mu _{1}}\nabla _{\mu
_{2}}R,\dots ,\nabla _{\mu _{1}} \dots \nabla _{\mu _{n}}R\right) $ theories of
gravity, are equivalent to scalar-multitensorial theories in the Jordan frame.
In particular, in the metric and Palatini formalisms, this
scalar-multitensorial equivalent scenario shows a structure that resembles
that of the Brans-Dicke theories with a kinetic term for the scalar field with $%
\omega _{0}=0$ or $\omega _{0}=-3/2$, respectively. In the present work, the
aforementioned analysis is extended to the Einstein frame. The conformal
transformation of the metric characterizing the transformation from Jordan's
to Einstein's frame is responsible for decoupling the scalar field from the
scalar curvature and also for introducing a usual kinetic term for the
scalar field in the metric formalism. In the Palatini approach, this
kinetic term is absent in the action. Concerning the other tensorial
auxiliary fields, they appear in the theory through a generalized potential.
As an example, the analysis of an extension of the Starobinsky model (with an
extra term proportional to $\nabla _{\mu }R\nabla ^{\mu }R$) is performed
and the fluid representation for the energy-momentum tensor is considered.
In the metric formalism, the presence of the extra term causes the fluid to
be an imperfect fluid with a heat flux contribution; on the other hand, in
the Palatini formalism the effective energy-momentum tensor for the extended
Starobinsky gravity is that of a perfect fluid type. Finally, it is
also shown that the extra term in the Palatini formalism represents a
dynamical field which is able to generate an inflationary regime without a
graceful exit.
\end{abstract}

\maketitle



\section{Introduction\label{sec:Introduction}}

The current precision-data era of observational cosmology \cite{PantheonSample2018,Planck2018,Planck2018Overview,Planck2018Parameters,DES2017} poses challenges for the standard model of particle physics and general relativity (GR) alike. In fact, the
particle physics community works intensely to accommodate dark matter (DM) within
the theoretical framework ---{} proposals include axionlike particles \cite%
{Peccei1977,Sasaki1984,Brandenberger2016,Marsh2016,Saikawa2017}, weakly interacting massive particles (WIMPs) \cite{Roszkowski2018}, superfluid DM \cite{Khoury2015,Brandenberger2018,Evan2018} ---{}
and experimental facilities strive to detect the dark matter particle \cite%
{Xenon2014,Lux2017}. Dark energy hints that general relativity may not be
the final theory of the gravitational interaction ---{} although it is
possible to explain it via a cosmological constant \cite{Carrol1992} or
exotic matter components. The last solution is the so-called modified matter
approach \cite{Amendola2010} whose particular models are: quintessence \cite%
{Caldwell1998,Carroll1998,Tsujikawa2013}, $k$-essence \cite%
{Chiba2000,Picon2001}, and unified models of dark matter and dark energy \cite%
{AstropartPhys2016,Cuzinatto2018,Brandenberger2019,Bento2002,Bertacca2010,Waga2003,Linder2005}%
. Dark energy could also be explained by the modified gravity approach, i.e.
extensions to GR changing the geometrical side of the field equation.
Probably, the most explored framework in this branch is the $f(R)$ theories
\cite{Cappozziello2008,SotiriouFaraoni2010,Nojiri2011,Vasilis2017} but
several other types have been explored \cite%
{Saridakis2016,Pablo2016,Jhingan2008,Shahin2007,Nojiri2003,Nojiri2007,Nojiri2008}%
. String-inspired theories \cite{Biswas2010,Biswas2012,Biswas2015} and
gauge-invariant gravity theories \cite{EPJC2008} suggest that a possible
suitable modification is to consider higher-order derivatives of the
curvature-related objects ($R$, $R_{\mu \nu }$, $R_{\,\nu \rho \sigma }^{\mu
}$); these theories will henceforth be called higher-order theories of
gravity. This possibility of extension to GR has drawn the attention of the
community as a possibility to address inflation \cite%
{Berkin1990,Gottlober1991,Amendola1993,Iihoshi2011,Diamandis2017,Koshelev2016,Koshelev2018,Edholm2017,Chialva2015}%
, efforts toward a meaningful quantization of gravity \cite%
{Modesto2012,Shapiro2015,ModestoShapiro2016,Ohta2018b} or the issue of ghost
plagued models \cite{Langlois2016a,Langlois2016b,Langlois2016c,Paul2017}
---{} see also \cite{Kaparulin2014}.

The interest in higher-order theories extends to the subject of their
equivalence to scalar-tensor theories. This is ultimately possible due to
the additional degrees of freedom (with respect to GR) that are present in both
approaches. The equivalence modified\textendash gravity/scalar-tensor
theories were studied both at the classical level \cite%
{PRD2016,SotiriouFaraoni2010} and at the quantum level \cite%
{Ruf2018a,Ruf2018b,Ohta2018a}, at least for $f\left(R\right)$ theories.

Higher-order $f\left( R,\nabla _{\mu _{1}}R,\nabla _{\mu _{1}}\nabla _{\mu
_{2}}R,\dots,\nabla _{\mu _{1}}\dots \nabla _{\mu _{n}}R\right) $ theories of
gravity are known to be scalar-multitensorial equivalent \cite{PRD2016}.
The equivalence was shown both in the metric formalism and in the Palatini
formalism in the Jordan frame. In the metric formalism, the scalar-multitensor
actions derived from $f\left( R,\nabla R,\dots,\nabla ^{n}R\right) $ are shown
to be analogous to the Brans-Dicke class of models with parameter $\omega
_{0}=0$ and a potential $U$; the theory differs from ordinary Brans-Dicke in
the sense that the degrees of freedom additional to the scalar one appear
within the definition of $U$ ---{} these additional components are tensorial
in character and motivate the name \textquotedblleft scalar-multitensorial
equivalent\textquotedblright . In the Palatini formalism, this equivalence
resembles a Brans-Dicke theory with $\omega _{0}=-3/2$ again with a
potential. In both formalisms, the equivalence was established in the Jordan
frame, where the scalar mode couples with the Ricci scalar in the action. This
action is said to be in the Einstein frame when a Ricci scalar is not
accompanied by any field (scalar or otherwise) and the extra fields appear
in the potential or even explicitly in the action integral (but not coupled
with curvature-related objects). A considerable advantage of the Einstein frame
is that the theory is rewritten as GR plus extra fields minimally coupled
with gravity. The Einstein frame is derived from the Jordan frame through a
conformal transformation of the metric tensor and convenient field
redefinitions of the scalar field and tensor fields eventually present in
the action. The passage from the Jordan to the Einstein frame is a step that is
missing in the work \cite{PRD2016}.\footnote{%
There is a way to transform the action from the original (geometric) frame
directly to the Einstein frame in $f(R)$ theories as explained in Refs.~\cite{BarrowCotsakis88,Witt84}. See also \cite{Orazi2017} for the Palatini formalism case.} One of the present paper's main goals is to fill in this gap and advance the study of higher-order $f\left( R,\nabla _{\mu _{1}}R,\dots,\nabla
_{\mu _{1}}\dots\nabla _{\mu _{n}}R\right) $ gravity.

The tools developed in this context will be applied here to a particular model of inflation. As it is well known from the literature, nowadays the Starobinsky model \cite{Starobinsky1980} is considered the most promising candidate for describing the inflationary period of the universe \cite{Planck2018}.\footnote{Examples of nonstandard proposals to the mechanism responsible for the early universe accelerated dynamics are presented in Refs.~\cite{Cu08EPJC,Cu08IJMPD,c-flation}.} This model modifies GR by adding a single term proportional to $R^{2}$ to the Einstein-Hilbert Lagrangian. For this reason, it is one of the most minimalist and simplest proposals for inflation. It is also motivated by the introduction of vacuum quantum corrections to the theory of gravity \cite{Zeldovich1972}. However, the main success of Starobinsky model lies in the fact that it is completely compatible with the most recent observational data \cite{Planck2018,Planck2018Overview,Planck2018Parameters}. Even though the current precision of the available data still cannot exclude other models as viable candidates. The scope here is to explore an extension of Starobinsky inflation stemming from the inclusion of the derivative-type term $\nabla_{\mu}R\nabla^{\mu}R$ alongside the Einstein-Hilbert term and an $R^{2}$ term in the gravitational Lagrangian. The resulting model was dubbed Starobinsky-Podolsky theory in Ref.~\cite{PRD2016}. This inflationary model has been fully explored in Ref.~\cite{StaPodInf} via the \emph{metric} formalism in the Einstein frame.\footnote{An equivalent model using two scalar auxiliary fields was analyzed in Ref.~\cite{Castellanos2018}. This paper assumes the higher-order term as a small perturbation to the Starobinsky Lagrangian.} Here its \emph{Palatini} counterpart shall be explored. There is a clear motivation for studying Starobinsky-Podolsky inflation in the Palatini approach. In fact, the Palatini version of the original Starobinsky model does not allow inflation to occur because the auxiliary scalar field in the theory is not a dynamical quantity \cite{SotiriouFaraoni2010}. We will show later in this paper that the Starobinsky-Podolsky model accommodates an auxiliary vector field besides the usual Starobinsky's traditional auxiliary scalar field. The presence of this additional vector field may change the dynamics of the system. The consequences for inflation of introducing such a higher-order term will be considered. In particular, the intention is to check if inflation is attainable within the Starobinsky-Podolsky model in the Palatini formalism and if a graceful exit can take place in this approach.

The paper is organized as follows: Initially, the Jordan-to-Einstein frame transformation is performed at the level of the action. Section \ref{sec:Einstein-frame} deals with the construction
of the scalar-multitensorial equivalent of $f\left( R,\nabla R,\dots,\nabla
^{n}R\right) $ in the Einstein frame in both metric and Palatini formalisms. The
Starobinsky-Podolsky theory \cite{PRD2016} is taken as a paradigm of
nonsingular $f\left( R,\nabla R\right) $ gravity in Sec.~\ref{sec:Starobinsky-Podolsky};
the general technique developed in Sec.~\ref{sec:Einstein-frame} is applied
to this case and gives rise to a natural definition of an effective energy-momentum
tensor $\tilde{T}_{\mu \nu}^{\left( \text{eff}\right) }$. The fluid representation of
$\tilde{T}_{\mu \nu }^{\left( \text{eff}\right) }$ is explored in Sec.~\ref{subsec:Fluid-Teff}. The 
Starobinsky-Podolsky theory in the context of inflation is
dealt with in Sec.~\ref{sec:Inflation}. Final comments are displayed in Sec.~\ref{sec:Final-remarks}.


\section{Higher-order gravity in the Einstein frame \label{sec:Einstein-frame}}

We start with a generic higher-order gravity theory of the form\footnote{%
Throughout the signature is $\left(-,+,+,+\right)$. We adopt units where $%
c=1 $ and $2\kappa=16\pi G=1$.}
\begin{equation}
S=\int d^{4}x\sqrt{-g}\left[f\left(R,\nabla
R,\nabla^{2}R,\dots,\nabla^{n}R\right)+\mathcal{L}_{M}\right].
\label{eq:S-geometric}
\end{equation}
This is called the geometric frame for the action integral. In Ref.~\cite{PRD2016},
we have shown it can be written in the Jordan frame
\begin{equation}
S^{\prime}=\int d^{4}x\sqrt{-g}\left[\Phi R-\omega_{0}\frac{1}{\Phi}%
\partial_{\rho}\Phi\partial^{\rho}\Phi-U\left(\Phi,\phi_{\mu},\dots,\phi_{%
\mu_{1} \dots \mu_{n}},\nabla_{\mu}\phi^{\mu},\dots,\nabla_{\mu_{n}}\dots\nabla_{%
\mu_{1}}\phi^{\mu_{1}\dots\mu_{n}}\right)+\mathcal{L}_{M}\left(g_{\mu\nu},%
\psi,\nabla\psi,\dots\right)\right],  \label{eq:S-Jordan}
\end{equation}
where $\psi$ are the matter fields. The value of the parameter $\omega_{0}$
distinguishes the formalism used to describe the theory: $\omega_{0}=0$ is
used for the metric formalism and $\omega_{0}=-\frac{3}{2}$ is a feature of
the Palatini one. In the latter, the action $S^{\prime}$ is required to
depend only on the Levi-Civita connection. In both Palatini and metric
frames, the use of auxiliary fields is required. These auxiliary fields
---{} appearing in Eq.~(\ref{eq:S-Jordan}) ---{} are defined
as
\begin{equation}
\Phi\equiv\phi-\nabla_{\mu}\phi^{\mu}+\dots+\left(-1\right)^{n}\nabla_{%
\mu_{n}}\dots\nabla_{\mu_{1}}\phi^{\mu_{1}\dots\mu_{n}},  \label{eq:Phi}
\end{equation}
\begin{equation}
\begin{cases}
\phi\equiv\frac{\partial f}{\partial\xi} \, , \\
\phi^{\mu}\equiv\frac{\partial f}{\partial\xi_{\mu}} \, , \\
\phi^{\mu\nu}\equiv\frac{\partial f}{\partial\xi_{\mu\nu}} \, ,\\
\vdots \\
\phi^{\mu_{1}\dots\mu_{n}}\equiv\frac{\partial f}{\partial\xi_{\mu_{1}\dots\mu_{n}}} \, ,
\end{cases}
\label{eq:Multi-Tensor}
\end{equation}
and
\begin{align}
U\left(\Phi,\phi_{\mu},\dots,\phi_{\mu_{1}\dots\mu_{n}},\nabla_{\mu}\phi^{%
\mu},\dots,\nabla_{\mu_{n}}\dots\nabla_{\mu_{1}}\phi^{\mu_{1}\dots\mu_{n}}\right)=
&
\left(\Phi+\nabla_{\mu}\phi^{\mu}+\dots+\left(-1\right)^{n+1}\nabla_{%
\mu_{n}}\dots\nabla_{\mu_{1}}\phi^{\mu_{1}\dots\mu_{n}}\right)\xi  \notag \\
&
+\phi^{\mu}\xi_{\mu}+\phi^{\mu\nu}\xi_{\mu\nu}+\dots+\phi^{\mu_{1}\dots\mu_{n}}%
\xi_{\mu_{1}\dots\mu_{n}}  \notag \\
& -f\left(\xi,\xi_{\mu},\xi_{\mu\nu},\dots,\xi_{\mu_{1}\dots\mu_{n}}\right).
\label{eq:U}
\end{align}
We indicate Ref. \cite{PRD2016} for further details.

In this paper, we proceed by introducing the conformal transformation for
the metric tensor
\begin{equation}
\tilde{g}_{\mu\nu}=\Phi g_{\mu\nu}  \label{eq:g-tilde}
\end{equation}
depending on the scalar field $\Phi$ given in Eq.~(\ref{eq:Phi}). Notice
that it depends on the contraction of the scalar and multitensor fields in
Eq.~(\ref{eq:Multi-Tensor}). Under Eq.~(\ref{eq:g-tilde}), the Ricci scalar
can be written as \cite{Wald84}\textbf{\ }
\begin{equation}
\tilde{R}\equiv\tilde{g}^{\mu\nu}\tilde{R}_{\mu\nu}=\frac{1}{\Phi}\left[R+%
\frac{3}{2\Phi^{2}}\nabla_{\rho}\Phi\nabla^{\rho}\Phi-\frac{3}{\Phi}%
\square\Phi\right]  \label{eq:R-tilde}
\end{equation}
or
\begin{equation}
R=\Phi\tilde{R}-\frac{3}{2\Phi}\tilde{\partial}_{\rho}\Phi\tilde{\partial}%
^{\rho}\Phi+\frac{3\Phi}{\sqrt{-\tilde{g}}}\tilde{\partial}_{\rho}\left(%
\sqrt{-\tilde{g}}\tilde{\partial}^{\rho}\ln\Phi\right),  \label{eq:R}
\end{equation}
since $\tilde{\partial}^{\rho}\equiv\tilde{g}^{\rho\alpha}\tilde{\partial}%
_{\alpha}=\tilde{g}^{\rho\alpha}\partial_{\alpha}$.

Equation (\ref{eq:R}) converts Eq.~(\ref{eq:S-Jordan}) to
\begin{align}
S^{\prime\prime}= & \int d^{4}x\sqrt{-\tilde{g}}\left\{ \tilde{R}-\frac{1}{%
\Phi^{2}}\left(\omega_{0}+\frac{3}{2}\right)\tilde{\partial}_{\rho}\Phi%
\tilde{\partial}^{\rho}\Phi+\frac{3}{\sqrt{-\tilde{g}}}\tilde{\partial}%
_{\rho}\left(\sqrt{-\tilde{g}}\tilde{\partial}^{\rho}\ln\Phi\right)\right.
\notag \\
& \left.-\frac{1}{\Phi^{2}}U\left(\Phi,\phi_{\mu},\dots,\phi_{\mu_{1}\dots%
\mu_{n}},\nabla_{\mu}\phi^{\mu},\dots,\nabla_{\mu_{n}}\dots\nabla_{\mu_{1}}%
\phi^{\mu_{1}\dots\mu_{n}}\right)+\frac{1}{\Phi^{2}}\mathcal{L}%
_{M}\left(g_{\mu\nu},\psi,\nabla\psi,\dots\right)\right\} .
\label{eq:S-Einstein}
\end{align}
This is the general form of the action in the Einstein frame. Next, we obtain
the field equations in both the metric and the Palatini formalisms.


\subsection{Metric formalism\label{subsec:Metric-formalism}}

Here, we will proceed by taking $\omega_{0}\neq-\frac{3}{2}$ which includes
the case $\omega_{0}=0$. We start by redefining the scalar field $\Phi$:
\begin{equation}
\tilde{\Phi}=\sqrt{3+2\omega_{0}}\ln\Phi.  \label{eq:Phi-tilde}
\end{equation}

The conformal transformation for the metric (\ref{eq:g-tilde}) implies
\begin{equation}
\Gamma_{\mu\rho}^{\alpha}=\tilde{\Gamma}_{\mu\rho}^{\alpha}+\Upsilon_{\rho%
\mu}^{\alpha},  \label{eq:Gamma-tilde}
\end{equation}
with the object $\Upsilon_{\rho\mu}^{\alpha}$ defined as
\begin{align}
\Upsilon_{\rho\mu}^{\nu} & =-\frac{1}{2}\left(\delta_{\rho}^{\nu}\partial_{%
\mu}\ln\Phi+\delta_{\mu}^{\nu}\partial_{\rho}\ln\Phi-\tilde{g}^{\nu\sigma}%
\tilde{g}_{\mu\rho}\partial_{\sigma}\ln\Phi\right)  \notag \\
& =-\frac{1}{2\sqrt{3}}\left(\delta_{\rho}^{\nu}\partial_{\mu}\tilde{\Phi}%
+\delta_{\mu}^{\nu}\partial_{\rho}\tilde{\Phi}-\tilde{g}^{\nu\sigma}\tilde{g}%
_{\mu\rho}\partial_{\sigma}\tilde{\Phi}\right)  \label{eq:T}
\end{align}
and $\tilde{\Gamma}_{\mu\rho}^{\alpha}$ is the Levi-Civita connection for
the metric $\tilde{g}_{\mu\nu}$. Henceforth, we will denote the operator $%
\tilde{\nabla}$ as the covariant derivative with respect to the connection $\tilde{%
\Gamma}$ and $\Upsilon_{\mu\rho}^{\sigma}=\Upsilon_{\mu\rho}^{\sigma}\left(%
\tilde{\Phi},\tilde{g}^{\alpha\beta}\right)$. With this, it is easy to show
the following useful formulas:
\begin{equation}
\nabla_{\mu}\phi^{\mu}=\tilde{\nabla}_{\mu}\phi^{\mu}+\Upsilon_{\rho\mu}^{%
\rho}\phi^{\mu}=\tilde{\nabla}_{\mu}\phi^{\mu}-\frac{2}{\sqrt{3}}%
\phi^{\rho}\partial_{\rho}\tilde{\Phi}  \label{eq:Nabla(Nabla-tilde)}
\end{equation}
and
\begin{equation}
\nabla_{\nu}\nabla_{\mu}\phi^{\mu\nu}=\tilde{\nabla}_{\nu}\left[\tilde{\nabla%
}_{\mu}\phi^{\mu\nu}+\Upsilon_{\mu\rho}^{\mu}\phi^{\rho\nu}+\Upsilon_{\mu%
\rho}^{\nu}\phi^{\mu\rho}\right]+\Upsilon_{\nu\sigma}^{\nu}\left(\tilde{%
\nabla}_{\mu}\phi^{\mu\sigma}+\Upsilon_{\mu\rho}^{\mu}\phi^{\rho\sigma}+%
\Upsilon_{\mu\rho}^{\sigma}\phi^{\mu\rho}\right).
\label{eq:Nabla-Nabla(Nabla-tilde)}
\end{equation}

The action $S^{\prime\prime}$, modulo surface terms, then reads
\begin{equation}
S^{\prime\prime}=\int d^{4}x\sqrt{-\tilde{g}}\left[\tilde{R}-\frac{1}{2}%
\tilde{g}^{\rho\sigma}\partial_{\rho}\tilde{\Phi}\partial_{\sigma}\tilde{\Phi%
}-\tilde{U}+\mathcal{\tilde{L}}_{M}\right]  , \label{eq:S-metric}
\end{equation}
where the conformal potential is
\begin{align}
\tilde{U} & =\tilde{U}\left(\tilde{\Phi},\tilde{\nabla}_{\mu}\tilde{\Phi}%
,\dots,\tilde{\nabla}_{\mu_{n}}\dots\tilde{\nabla}_{\mu_{1}}\tilde{\Phi}%
,\phi^{\mu_{1}},\dots,\phi^{\mu_{1}\dots\mu_{n}},\tilde{\nabla}%
_{\mu_{1}}\phi^{\nu_{1}},\dots,\tilde{\nabla}_{\mu_{1}}\phi^{\nu_{1}\dots%
\nu_{n}},\right.  \notag \\
& \left.\qquad\tilde{\nabla}_{\mu_{1}}\tilde{\nabla}_{\mu_{2}}\phi^{\nu_{1}%
\nu_{2}},\dots,\tilde{\nabla}_{\mu_{1}}\tilde{\nabla}_{\mu_{2}}\phi^{%
\nu_{1}\dots\nu_{n}},\dots,\tilde{\nabla}_{\mu_{1}}\dots\tilde{\nabla}%
_{\mu_{n}}\phi^{\nu_{1}\dots\nu_{n}}\right)  \notag \\
& =e^{-\frac{2\tilde{\Phi}}{\sqrt{3+2\omega_{0}}}}U\left(\Phi\left(\tilde{%
\Phi}\right),\phi_{\mu},\dots,\phi_{\mu_{1}\dots\mu_{n}},\nabla_{\mu}\phi^{\mu}%
\left(\tilde{\Phi},\tilde{g}^{\alpha\beta}\right),\dots,\nabla_{\mu_{n}}\dots%
\nabla_{\mu_{1}}\phi^{\mu_{1}\dots\mu_{n}}\left(\tilde{\Phi},\tilde{g}%
^{\alpha\beta}\right)\right),  \label{eq:U-tilde}
\end{align}
and
\begin{equation}
\mathcal{\tilde{L}}_{M}=\mathcal{\tilde{L}}_{M}\left(\tilde{g}_{\mu\nu},%
\tilde{\Phi},\psi,\tilde{\nabla}\psi,\dots\right)=e^{-\frac{2\tilde{\Phi}}{%
\sqrt{3+2\omega_{0}}}}\mathcal{L}_{M}\left(g_{\mu\nu}\left(\tilde{g}%
_{\mu\nu},\tilde{\Phi}\right),\psi,\nabla\psi\left(\tilde{\Phi},\tilde{g}%
^{\alpha\beta}\right),\dots\right)  \label{eq:L_M-tilde}
\end{equation}
is the matter-field Lagrangian under the metric conformal transformation.

Equation (\ref{eq:S-metric}) is the action integral in the Einstein frame for the
metric formalism. The corresponding field equations are derived next.


\subsubsection{Field equations in the metric formalism\label%
{subsec:EOM-metric}}

Functional variations of the action $S^{\prime\prime}$ in Eq.~(\ref{eq:S-metric}) with respect to the fields give the equations of motion (EOM) for the higher-order gravity in the Einstein frame. In the metric approach,
these fields are the conformal metric $\tilde{g}^{\mu\nu}$, the scalar field
$\tilde{\Phi}$, the multitensor fields $\left\{\phi^{\mu_{1}},\phi^{\mu_{1}\mu_{2}},\dots,\phi^{\mu_{1}\dots\mu_{n}}\right\} $ and the matter field $\psi$. Performing the variations in this sequence leads
to the EOM Eqs.~(\ref{eq:EOM-metric-g}), (\ref{eq:EOM-metric-(Phi-tilde)}), 
(\ref{eq:EOM-metric-(multi-phi)}), and (\ref{eq:EOM-metric-psi})
below. In fact, the gravitational field equation is
\begin{equation}
\tilde{R}_{\mu\nu}-\frac{1}{2}\tilde{g}_{\mu\nu}\tilde{R}=\frac{1}{2}%
\left(T_{\mu\nu}+T_{\mu\nu}^{\left(\text{eff}\right)}\right),
\label{eq:EOM-metric-g}
\end{equation}
where
\begin{equation}
T_{\mu\nu}\equiv\tilde{g}_{\mu\nu}\mathcal{\tilde{L}}_{M}-2\frac{\delta%
\mathcal{\tilde{L}}_{M}}{\delta\tilde{g}^{\mu\nu}}  \label{eq:Tmunu}
\end{equation}
is the ordinary energy-momentum tensor for the matter field and
\begin{equation}
T_{\mu\nu}^{\text{(eff)}}\equiv\partial_{\mu}\tilde{\Phi}\partial_{\nu}%
\tilde{\Phi}-\tilde{g}_{\mu\nu}\left(\frac{1}{2}\tilde{g}^{\rho\sigma}%
\partial_{\rho}\tilde{\Phi}\partial_{\sigma}\tilde{\Phi}+\tilde{U}\right)+2%
\frac{\delta\tilde{U}}{\delta\tilde{g}^{\mu\nu}}  \label{eq:Tmunu-eff-metric}
\end{equation}
is the effective energy-momentum tensor for the auxiliary fields in the
metric formalism. The definition Eq.~(\ref{eq:Tmunu-eff-metric}) contains the
term
\begin{equation}
\frac{\delta\tilde{U}}{\delta\tilde{g}^{\mu\nu}}=\frac{\partial\tilde{U}}{%
\partial\tilde{g}^{\mu\nu}}-\frac{1}{\sqrt{-\tilde{g}}}\partial_{\mu_{1}}%
\left(\sqrt{-\tilde{g}}\frac{\partial\tilde{U}}{\partial\left(\partial_{%
\mu_{1}}\tilde{g}^{\mu\nu}\right)}\right)+\dots+\frac{\left(-1\right)^{n}}{%
\sqrt{-\tilde{g}}}\partial_{\mu_{1}}\dots\partial_{\mu_{n-1}}\partial_{%
\mu_{n}}\left(\sqrt{-\tilde{g}}\frac{\partial\tilde{U}}{\partial\left(%
\partial_{\mu_{n}}\dots\partial_{\mu_{1}}\tilde{g}^{\mu\nu}\right)}\right) .
\label{eq:d(U-tilde)d(g-tilde)}
\end{equation}

Moreover, the EOM for the scalar field $\tilde{\Phi}$ is
\begin{equation}
\tilde{\square}\tilde{\Phi}-\frac{\delta\tilde{U}}{\delta\tilde{\Phi}}=-%
\frac{\delta\mathcal{\tilde{L}}_{M}}{\delta\tilde{\Phi}},
\label{eq:EOM-metric-(Phi-tilde)}
\end{equation}
where $\tilde{\square}=\tilde{\nabla}_{\rho}\tilde{\nabla}^{\rho}$, under
the definition
\begin{equation}
\frac{\delta\tilde{U}}{\delta\tilde{\Phi}}=\frac{\partial\tilde{U}}{\partial%
\tilde{\Phi}}-\tilde{\nabla}_{\mu_{1}}\frac{\partial\tilde{U}}{\partial\left(%
\tilde{\nabla}_{\mu_{1}}\tilde{\Phi}\right)}+\dots+\left(-1\right)^{n}\tilde{%
\nabla}_{\mu_{1}}\dots\tilde{\nabla}_{\mu_{n}}\frac{\partial\tilde{U}}{%
\partial\left(\tilde{\nabla}_{\mu_{n}}\dots\tilde{\nabla}_{\mu_{1}}\tilde{\Phi}%
\right)}.  \label{eq:d(U-tilde)d(Phi-tilde)}
\end{equation}

Variations with respect to the multiple tensorial fields $\left\{
\phi^{\mu_{1}},\phi^{\mu_{1}\mu_{2}},\dots,\phi^{\mu_{1}\dots\mu_{n}}\right\} $
lead to the set of equations
\begin{equation}
\begin{cases}
\frac{\partial\tilde{U}}{\partial\phi^{\nu_{1}}}-\tilde{\nabla}%
_{\mu_{1}}\left(\frac{\partial\tilde{U}}{\partial\tilde{\nabla}%
_{\mu_{1}}\phi^{\nu_{1}}}\right)=0 \, ,\\
\frac{\partial\tilde{U}}{\partial\phi^{\nu_{1}\nu_{2}}}-\tilde{\nabla}%
_{\mu_{1}}\left(\frac{\partial\tilde{U}}{\partial\tilde{\nabla}%
_{\mu_{1}}\phi^{\nu_{1}\nu_{2}}}\right)+\tilde{\nabla}_{\mu_{2}}\tilde{\nabla%
}_{\mu_{1}}\left(\frac{\partial\tilde{U}}{\partial\tilde{\nabla}_{\mu_{1}}%
\tilde{\nabla}_{\mu_{2}}\phi^{\nu_{1}\nu_{2}}}\right)=0 \, ,\\
\vdots \\
\frac{\partial\tilde{U}}{\partial\phi^{\nu_{1}\dots\nu_{n}}}-\tilde{\nabla}%
_{\mu_{1}}\left(\frac{\partial\tilde{U}}{\partial\tilde{\nabla}%
_{\mu_{1}}\phi^{\nu_{1}\dots\nu_{n}}}\right)+\tilde{\nabla}_{\mu_{2}}\tilde{%
\nabla}_{\mu_{1}}\left(\frac{\partial\tilde{U}}{\partial\tilde{\nabla}%
_{\mu_{1}}\tilde{\nabla}_{\mu_{2}}\phi^{\nu_{1}\dots\nu_{n}}}%
\right)-\dots+\left(-1\right)^{n}\tilde{\nabla}_{\mu_{n}}\dots\tilde{\nabla}%
_{\mu_{1}}\left(\frac{\partial\tilde{U}}{\partial\tilde{\nabla}_{\mu_{1}}\dots%
\tilde{\nabla}_{\mu_{n}}\phi^{\nu_{1}\dots\nu_{n}}}\right)=0 \, .%
\end{cases}
\label{eq:EOM-metric-(multi-phi)}
\end{equation}

Finally, the EOM for the matter field $\psi$ reads simply
\begin{equation}
\frac{\delta\mathcal{\tilde{L}}_{M}}{\delta\psi}=0\text{.}
\label{eq:EOM-metric-psi}
\end{equation}

We end this section by emphasizing that $\tilde{U}$ depends on higher-order
derivatives of the fields $\tilde{\Phi}$ and $\phi^{\nu_{1}\dots\nu_{n}}$ (for
$n\geq2$); cf. Eq.~(\ref{eq:U-tilde}). These higher-order terms give rise to
additional kinetic terms in the scalar-tensor theory.


\subsection{Palatini formalism\label{subsec:Palatini-formalism}}

In the Palatini formalism, the metric tensor and the connection are
varied independently. This demands a different notation from the previous
metric formalism. Accordingly, we adopt
\begin{equation}
\mathcal{R}=g^{\mu\nu}\mathcal{R}_{\mu\nu}=g^{\mu\nu}\left(\partial_{\rho}%
\Gamma_{\mu\nu}^{\rho}-\partial_{\mu}\Gamma_{\rho\nu}^{\rho}+\Gamma_{\mu%
\nu}^{\beta}\Gamma_{\rho\beta}^{\rho}-\Gamma_{\rho\nu}^{\beta}\Gamma_{\mu%
\beta}^{\rho}\right).  \label{eq:R-Pal}
\end{equation}
If GR is required to be a limit case, then the covariant derivative $%
\nabla_{\rho}$ should be given in terms of the Christoffel symbols $\left\{
_{\rho\sigma}^{\tau}\right\} =\frac{1}{2}g^{\tau\lambda}\left(\partial_{%
\sigma}g_{\lambda\rho}+\partial_{\rho}g_{\sigma\lambda}-\partial_{%
\lambda}g_{\rho\sigma}\right)$. Moreover, the Ricci scalar is $%
R=g^{\mu\nu}R_{\mu\nu}=g^{\mu\nu}\left(\partial_{\rho}\left\{
_{\mu\nu}^{\rho}\right\} -\partial_{\mu}\left\{ _{\rho\nu}^{\rho}\right\}
+\left\{ _{\mu\nu}^{\beta}\right\} \left\{ _{\rho\beta}^{\rho}\right\}
-\left\{ _{\rho\nu}^{\beta}\right\} \left\{ _{\mu\beta}^{\rho}\right\}
\right)$. On the other hand, the covariant derivative constructed from $%
\Gamma$ is denoted by $\bar{\nabla}_{\rho}$. The action must contain
covariant derivatives built only with $\left\{ _{\rho\sigma}^{\tau}\right\} $%
; this is related to the way matter responds to gravity, i.e. $%
\nabla_{\mu}T^{\mu\nu}=0$ (but $\bar{\nabla}_{\mu}T^{\mu\nu}\neq0$) \cite%
{SotiriouFaraoni2010} ---{} see also \cite{Will81}.

The metric conformal to $g_{\mu\nu}$ defined as
\begin{equation}
h_{\mu\nu}\equiv f^{\prime}\left(\mathcal{R},\nabla\mathcal{R}%
,\dots\right)g_{\mu\nu}\,,  \label{eq:hmunu}
\end{equation}
involves the general derivative
\begin{align}
f^{\prime}\left(\mathcal{R},\nabla\mathcal{R},\dots\right) & =\frac{\partial f%
}{\partial\mathcal{R}}-\nabla_{\rho}\left(\frac{\partial f}{%
\partial\nabla_{\rho}\mathcal{R}}\right)+\dots  \notag \\
& \quad+\left(-1\right)^{n}\nabla_{\rho_{n}}\dots\nabla_{\rho_{1}}\frac{%
\partial f}{\partial\nabla_{\rho_{1}}\dots\nabla_{\rho_{n}}\mathcal{R}}\,.
\label{eq:f-prime}
\end{align}

The metric $h_{\mu\nu}$ satisfies
\begin{equation}
h^{\alpha\beta}=\frac{1}{f^{\prime}\left(\mathcal{R},\nabla\mathcal{R}%
,\dots\right)}g^{\alpha\beta}\,,  \label{eq:hmunu-contra}
\end{equation}
and
\begin{equation}
\sqrt{-h}=\left[f^{\prime}\left(\mathcal{R},\nabla\mathcal{R},\dots\right)%
\right]^{2}\sqrt{-g},\quad\left(h\equiv\det h_{\mu\nu}\right)\,,
\label{eq:h}
\end{equation}
so the metricity condition holds,
\begin{equation}
\bar{\nabla}_{\rho}\left(\sqrt{-h}h^{\alpha\beta}\right)=0\Rightarrow\bar{%
\nabla}_{\rho}h_{\alpha\beta}=0 \, .  \label{eq:h-metricity}
\end{equation}
Then a relation between $\Gamma_{\mu\nu}^{\beta}$ and $\left\{
_{\mu\nu}^{\beta}\right\} $ is achieved:
\begin{equation}
\Gamma_{\mu\nu}^{\beta}=\left\{ _{\mu\nu}^{\beta}\right\} +\frac{1}{2}\frac{1%
}{f^{\prime}}g^{\alpha\beta}\left(g_{\alpha\mu}\partial_{\nu}f^{\prime}+g_{%
\nu\alpha}\partial_{\mu}f^{\prime}-g_{\mu\nu}\partial_{\alpha}f^{\prime}%
\right).  \label{eq:Gamma(Christoffel)}
\end{equation}

In the face of that, $\mathcal{R}_{\mu\nu}$ is written in terms of Ricci
tensor $R_{\mu\nu}$ (and derivatives of $f^{\prime}$) as
\begin{align}
\mathcal{R}_{\mu\nu} & =R_{\mu\nu}+\frac{3}{2}\frac{1}{\left(f^{\prime}%
\right)^{2}}\nabla_{\mu}f^{\prime}\nabla_{\nu}f^{\prime}  \notag \\
& \quad-\frac{1}{2}\frac{1}{f^{\prime}}\left(\nabla_{\mu}\nabla_{\nu}f^{%
\prime}+\nabla_{\nu}\nabla_{\mu}f^{\prime}+g_{\mu\nu}\square
f^{\prime}\right).  \label{eq:Rmunu(Ricci)}
\end{align}
For the scalar curvature
\begin{equation}
\mathcal{R}=R+\frac{3}{2}\frac{1}{\left(f^{\prime}\right)^{2}}%
\left(\nabla_{\mu}f^{\prime}\nabla^{\mu}f^{\prime}\right)-3\frac{1}{%
f^{\prime}}\left(\square f^{\prime}\right).  \label{eq:R(ScalarCurvature)}
\end{equation}

In the Palatini approach, Eq.~(\ref{eq:S-geometric}) is more clearly written
in the form
\begin{equation}
S=\int d^{4}x\sqrt{-g}\left[f\left(\mathcal{R},\nabla\mathcal{R},\nabla^{2}%
\mathcal{R},\dots,\nabla^{n}\mathcal{R}\right)+\mathcal{L}_{M}\right].
\label{eq:S-geometric-Pal}
\end{equation}
Under the definitions Eq.~(\ref{eq:Phi}), (\ref{eq:Multi-Tensor}), and
(\ref{eq:U}), the Jordan frame arises:
\begin{equation}
S^{\prime}=\int d^{4}x\sqrt{-g}\left[\Phi\mathcal{R}-U\left(\Phi,\phi_{%
\mu},\dots,\phi_{\mu_{1}\dots\mu_{n}},\nabla_{\mu}\phi^{\mu},\dots,\nabla_{%
\mu_{n}}\dots\nabla_{\mu_{1}}\phi^{\mu_{1}\dots\mu_{n}}\right)+\mathcal{L}%
_{M}\left(g_{\mu\nu},\psi,\nabla\psi,\dots\right)\right].
\label{eq:S-Jordan-Pal}
\end{equation}
The use of Eq.~(\ref{eq:R(ScalarCurvature)}) then leads to
\begin{equation*}
S^{\prime}=\int d^{4}x\sqrt{-g}\left[\Phi R+\frac{3}{2}\frac{1}{\Phi}%
\partial_{\rho}\Phi\partial^{\rho}\Phi-U\left(\Phi,\phi_{\mu},\dots,\phi_{%
\mu_{1}\dots\mu_{n}},\nabla_{\mu}\phi^{\mu},\dots,\nabla_{\mu_{n}}\dots\nabla_{%
\mu_{1}}\phi^{\mu_{1}\dots\mu_{n}}\right)+\mathcal{L}_{M}\left(g_{\mu\nu},%
\psi,\nabla\psi,\dots\right)\right],
\end{equation*}
which is precisely Eq.~(\ref{eq:S-Jordan}) with $\omega_{0}=-\frac{3}{2}$.
This justifies our statement below that equation. Therefore, the conformal
transformation $g_{\mu\nu}\rightarrow\tilde{g}_{\mu\nu}$ as in Eq.~(\ref%
{eq:g-tilde}) set the action integral in the Einstein frame:
\begin{equation}
S^{\prime\prime}=\int d^{4}x\sqrt{-\tilde{g}}\left[\tilde{R}-\tilde{U}+%
\mathcal{\tilde{L}}_{M}\left(\tilde{g}_{\mu\nu},\psi,\tilde{\nabla}%
\psi,\dots\right)\right].  \label{eq:S-Einstein-Pal}
\end{equation}
This is Eq.~(\ref{eq:S-Einstein}) with $\omega_{0}=-\frac{3}{2}$ and the
further definitions
\begin{align}
\tilde{U} & =\tilde{U}\left(\Phi,\tilde{\nabla}_{\mu}\Phi,\dots,\tilde{\nabla}%
_{\mu_{n}}\dots\tilde{\nabla}_{\mu_{1}}\Phi,\phi^{\mu_{1}},\dots,\phi^{%
\mu_{1}\dots\mu_{n}},\tilde{\nabla}_{\mu_{1}}\phi^{\nu_{1}},\dots,\tilde{\nabla}%
_{\mu_{1}}\phi^{\nu_{1}\dots\nu_{n}},\right.  \notag \\
& \qquad\left.\tilde{\nabla}_{\mu_{1}}\tilde{\nabla}_{\mu_{2}}\phi^{\nu_{1}%
\nu_{2}},\dots,\tilde{\nabla}_{\mu_{1}}\tilde{\nabla}_{\mu_{2}}\phi^{%
\nu_{1}\dots\nu_{n}},\dots,\tilde{\nabla}_{\mu_{1}}\dots\tilde{\nabla}%
_{\mu_{n}}\phi^{\nu_{1}\dots\nu_{n}}\right)  \notag \\
& =\frac{1}{\Phi^{2}}U\left(\Phi,\phi_{\mu},\dots,\phi_{\mu_{1}\dots\mu_{n}},%
\nabla_{\mu}\phi^{\mu}\left(\Phi,\tilde{g}^{\alpha\beta}\right),\dots,\nabla_{%
\mu_{n}}\dots\nabla_{\mu_{1}}\phi^{\mu_{1}\dots\mu_{n}}\left(\Phi,\tilde{g}%
^{\alpha\beta}\right)\right),  \label{eq:U-tilde-Pal}
\end{align}
and
\begin{equation}
\mathcal{\tilde{L}}_{M}\left(\Phi,\tilde{g}_{\mu\nu},\psi,\tilde{\nabla}%
\psi,\dots\right)=\frac{1}{\Phi^{2}}\mathcal{L}_{M}\left(g_{\mu\nu}\left(%
\tilde{g}_{\mu\nu},\Phi\right),\psi,\nabla\psi\left(\Phi,\tilde{g}%
^{\alpha\beta}\right),\dots\right),  \label{eq:L_M-tilde-Pal}
\end{equation}
with the ``conformal\textquotedblright{} covariant derivative $\tilde{\nabla}%
_{\mu}$ built from
\begin{equation}
\tilde{\Gamma}_{\mu\rho}^{\alpha}\left(\Phi,\tilde{g}^{\alpha\beta}\right)=%
\Gamma_{\mu\rho}^{\alpha}\left(g^{\alpha\beta}\left(\Phi,\tilde{g}%
^{\alpha\beta}\right)\right)-\Upsilon_{\rho\mu}^{\alpha}\left(\Phi,\tilde{g}%
^{\alpha\beta}\right) , \label{eq:Gamma-tilde-Pal}
\end{equation}
where
\begin{equation}
\Upsilon_{\rho\mu}^{\nu}\left(\Phi,\tilde{g}^{\alpha\beta}\right)=-\frac{1}{2%
}\left(\delta_{\rho}^{\nu}\partial_{\mu}\ln\Phi+\delta_{\mu}^{\nu}\partial_{%
\rho}\ln\Phi-\tilde{g}^{\nu\sigma}\tilde{g}_{\mu\rho}\partial_{\sigma}\ln%
\Phi\right).  \label{eq:T-Pal}
\end{equation}
Compare Eqs.~(\ref{eq:U-tilde-Pal}) and (\ref{eq:L_M-tilde-Pal}) to Eqs.~(%
\ref{eq:U-tilde}) and (\ref{eq:L_M-tilde}) ---{} notice that it
is not necessary to introduce the auxiliary field $\tilde{\Phi}$, as in Eq.~(%
\ref{eq:Phi-tilde}), since there is no kinetic term in $S^{\prime\prime}$
this time.


\subsubsection{Field equations in the Palatini formalism\label%
{subsec:EOM-Palatini}}

In accordance with the previous paragraph, the action $S^{\prime\prime}$ in
Eq.~(\ref{eq:S-Einstein-Pal}) does not depend explicitly on the general
connection $\Gamma$. Therefore, the field equations can be obtained by taking
variations with respect to the fields $\tilde{g}^{\mu\nu}$, $\Phi$, $%
\phi^{\mu}$, and so on.

The EOM for the gravitational field is formally the
same as Eq.~(\ref{eq:EOM-metric-g}) in the metric approach with the usual
energy-momentum tensor identical to Eq.~(\ref{eq:Tmunu}), but with an
effective energy-momentum tensor
\begin{equation}
T_{\mu\nu}^{\left(\text{eff}\right)}\equiv-\left(\tilde{g}_{\mu\nu}\tilde{U}%
-2\frac{\delta\tilde{U}}{\delta\tilde{g}^{\mu\nu}}\right),
\label{eq:Tmunu-eff-Pal}
\end{equation}
which differs from Eq.~(\ref{eq:Tmunu-eff-metric}). The second term in the
right-hand side (RHS) of Eq.~(\ref{eq:Tmunu-eff-Pal}) is given by Eq.~(\ref%
{eq:d(U-tilde)d(g-tilde)}). It is interesting to note that the potential $%
\tilde{U}$ gives rise to an effective energy-momentum tensor similar in
structure to the matter one up to a global sign.

Varying Eq.~(\ref{eq:S-Einstein-Pal}) with respect to $\Phi$ leads to
\begin{equation}
\frac{\delta\tilde{U}}{\delta\Phi}=\frac{\delta\mathcal{\tilde{L}}_{M}}{%
\delta\Phi},  \label{eq:EOM-Pal-Phi}
\end{equation}
where $\frac{\delta\tilde{U}}{\delta\Phi}$ is completely analogous to Eq.~(%
\ref{eq:d(U-tilde)d(Phi-tilde)}) albeit $\Phi$ is used in place of $\tilde{%
\Phi}$. When we confront Eq.~(\ref{eq:EOM-Pal-Phi}) with Eq.~(\ref%
{eq:EOM-metric-(Phi-tilde)}), we notice that a term of the type $\tilde{%
\square}\Phi$ is absent in the former, whereas it is present in the last.

The EOM for the multiple tensorial fields $\left\{
\phi^{\mu_{1}},\phi^{\mu_{1}\mu_{2}},\dots,\phi^{\mu_{1}\dots\mu_{n}}\right\} $
and the matter field are identical in form to Eqs.~(\ref%
{eq:EOM-metric-(multi-phi)}) and (\ref{eq:EOM-metric-psi}).

The next section deals with a case study: a particular Lagrangian of the
type $f\left(R,\nabla R\right)$ scaling with the Einstein-Hilbert term, the
Starobinsky contribution $R^{2}$, and a derivative term of the kind $%
\nabla_{\mu}R\nabla^{\mu}R$. We call this example the Starobinsky-Podolsky
gravity.


\section{Application: the Starobinsky-Podolsky Lagrangian\label%
{sec:Starobinsky-Podolsky}}

The original Starobinsky-Podolsky action is
\begin{equation}
S=\int d^{4}x\sqrt{-g}\left[R+\frac{c_{0}}{2}R^{2}+\frac{c_{1}}{2}%
\nabla_{\mu}R\nabla^{\mu}R\right],  \label{eq:S-SP-geometric}
\end{equation}
which in Jordan frame reads
\begin{equation}
S^{\prime}=\int d^{4}x\sqrt{-g}\left[\Phi R-\frac{1}{2c_{0}}%
\left(\Phi+\nabla_{\mu}\phi^{\mu}-1\right)^{2}-\frac{\phi^{\mu}\phi_{\mu}}{%
2c_{1}}\right],  \label{eq:S-SP-Jordan}
\end{equation}
as detailed shown in Ref.~\cite{PRD2016}. A version of Eq.~(\ref{eq:S-SP-geometric}) with $c_0 = 0$ was introduced by Ref.~\cite{EPJC2008} in the context of a second order gauge theory for gravity \cite{Cuzinatto:2005zr}. Some aspects of such a model have been investigated in Refs.~\cite{Cuzinatto:2007hf,Cuzinatto:2013pva} with respect to the present day acceleration.

Rigorously, Eqs.~(\ref{eq:S-SP-geometric}) and (\ref{eq:S-SP-Jordan}) have a
notation compatible with the metric formalism. For the Palatini
formalism, the mapping $R\rightarrow\mathcal{R}$ is required as explained in
the beginning of Sec.~\ref{subsec:Palatini-formalism}. Due to this
difference, we split the analysis in the two cases discussed in Sec.~\ref%
{subsec:SP-metric} and \ref{subsec:SP-Palatini}.


\subsection{Metric formalism\label{subsec:SP-metric}}

Equation (\ref{eq:S-SP-Jordan}) can be transformed using the definition Eq.~(\ref%
{eq:g-tilde}) of the conformal metric $\tilde{g}_{\mu\nu}$ and Eq.~(\ref%
{eq:Phi-tilde}) which introduced the field $\tilde{\Phi}$. The
Einstein-frame version for Starobinsky-Podolsky action then follows
\begin{equation}
S^{\prime\prime}=\int d^{4}x\sqrt{-\tilde{g}}\left[\tilde{R}-\frac{1}{2}%
\tilde{\partial}_{\rho}\tilde{\Phi}\tilde{\partial}^{\rho}\tilde{\Phi}-%
\tilde{U}\right],  \label{eq:S-SP-Einstein}
\end{equation}
where
\begin{equation}
\tilde{U}=\frac{1}{2c_{0}}\left(1+e^{-\frac{\tilde{\Phi}}{\sqrt{3}}}\tilde{%
\nabla}_{\mu}\phi^{\mu}-\frac{2e^{-\frac{\tilde{\Phi}}{\sqrt{3}}}}{\sqrt{3}}%
\phi^{\rho}\tilde{\partial}_{\rho}\tilde{\Phi}-e^{-\frac{\tilde{\Phi}}{\sqrt{%
3}}}\right)^{2}+e^{-\frac{3\tilde{\Phi}}{\sqrt{3}}}\frac{\tilde{g}%
_{\mu\nu}\phi^{\mu}\phi^{\nu}}{2c_{1}}.  \label{eq:U-tilde-metric-SP}
\end{equation}

It is worth mentioning that Eq.~(\ref{eq:S-SP-Einstein}) reproduces the
conventional Starobinsky Lagrangian \cite{Starobinsky1980} with the
potential
\begin{equation}
\tilde{U}=U\left(\tilde{\Phi}\right)=\frac{1}{2c_{0}}\left(1-e^{-\frac{%
\tilde{\Phi}}{\sqrt{3}}}\right)^{2} , \label{eq:U-S}
\end{equation}
by assuming $\phi^{\mu}=0$ in Eq.~(\ref{eq:U-tilde-metric-SP}).

The field equations for $\Phi$ and $\phi^{\mu}$ are
\begin{equation}
\tilde{\square}\tilde{\Phi}+\frac{1}{\sqrt{3}c_{0}}e^{-\frac{2\tilde{\Phi}}{%
\sqrt{3}}}\left(\frac{2}{\sqrt{3}}\phi^{\sigma}\tilde{\partial}_{\sigma}%
\tilde{\Phi}-\tilde{\nabla}_{\mu}\phi^{\mu}-1\right)\left(e^{\frac{\tilde{%
\Phi}}{\sqrt{3}}}+\tilde{\nabla}_{\mu}\phi^{\mu}-\frac{2}{\sqrt{3}}%
\phi^{\rho}\tilde{\partial}_{\rho}\tilde{\Phi}-1\right)-\frac{1}{\sqrt{3}}%
e^{-\frac{3\tilde{\Phi}}{\sqrt{3}}}\frac{\tilde{g}_{\mu\nu}\phi^{\mu}\phi^{%
\nu}}{2c_{1}}=0,  \label{eq:EOM-metric-(Phi-tilde)-SP}
\end{equation}
where $\tilde{\square}=\tilde{\nabla}_{\rho}\tilde{\nabla}^{\rho}$ and
\begin{equation}
\frac{1}{c_{0}}\tilde{\partial}_{\sigma}\left(e^{\frac{\tilde{\Phi}}{\sqrt{3}%
}}+\tilde{\nabla}_{\mu}\phi^{\mu}-\frac{2}{\sqrt{3}}\phi^{\mu}\tilde{\partial%
}_{\mu}\tilde{\Phi}-1\right)-\frac{1}{c_{1}}e^{-\frac{\tilde{\Phi}}{\sqrt{3}}%
}\phi_{\sigma}=0 . \label{eq:EOM-metric-(phi-mu)-SP}
\end{equation}
From Eqs.~(\ref{eq:EOM-metric-(Phi-tilde)-SP}) and (\ref%
{eq:EOM-metric-(phi-mu)-SP}), we conclude that both $\tilde{\Phi}$ and $%
\phi^{\mu}$ are dynamical fields in the sense of their Cauchy data. In
particular, due to the quadratic coupling $\left(\phi^{\rho}\partial_{\rho}%
\tilde{\Phi}\right)^{2}$ the second-order time derivative of $\tilde{\Phi}$
is present in both equations. With regard to the second-order derivative of $%
\phi^{\mu}$, only the $\phi^{0}$ component is derived twice with respect to
time ---{} this happens for the choice $\sigma=0$ in Eq.~(\ref%
{eq:EOM-metric-(phi-mu)-SP}). Therefore, only $\phi^{0}$ is dynamical, while
the equations for $\sigma=i$ establish constraints for the components $%
\phi^{i}$.

Predictably, the gravitational EOM for the Starobinsky-Podolsky Lagrangian in
the metric approach is\footnote{The ordinary $T_{\mu\nu}$ as given in Eq.~(\ref{eq:Tmunu}) is not present in Eq.~(\ref{eq:EOM-metric-g-SP}) because we are not taking the matter field in the analysis of Starobinsky-Podolsky action. This could simply be added to the theory afterwards.}
\begin{equation}
\tilde{R}_{\mu\nu}-\frac{1}{2}\tilde{g}_{\mu\nu}\tilde{R}=\frac{1}{2}%
T_{\mu\nu}^{\left(\text{eff}\right)},  \label{eq:EOM-metric-g-SP}
\end{equation}
where
\begin{equation}
\tilde{T}_{\mu\nu}^{\left(\text{eff}\right)}=\tilde{\partial}_{\mu}\tilde{%
\Phi}\tilde{\partial}_{\nu}\tilde{\Phi}-\tilde{g}_{\mu\nu}\left(\frac{1}{2}%
\tilde{g}^{\rho\alpha}\tilde{\partial}_{\rho}\tilde{\Phi}\tilde{\partial}%
_{\alpha}\tilde{\Phi}+\tilde{U}\right)+2\frac{\delta\tilde{U}}{\delta\tilde{g%
}^{\mu\nu}},  \label{eq:Tmunu-eff-metric-SP}
\end{equation}
with
\begin{equation}
2\frac{\delta\tilde{U}}{\delta\tilde{g}^{\mu\nu}}=-\frac{e^{-\frac{3\tilde{%
\Phi}}{\sqrt{3}}}}{c_{1}}\phi_{\mu}\phi_{\nu}+\frac{1}{c_{0}}\tilde{g}%
_{\mu\nu}\tilde{\nabla}_{\lambda}\left[\phi^{\lambda}e^{-\frac{\tilde{\Phi}}{%
\sqrt{3}}}\left(1+e^{-\frac{\tilde{\Phi}}{\sqrt{3}}}\tilde{\nabla}%
_{\rho}\phi^{\rho}-\frac{2e^{-\frac{\tilde{\Phi}}{\sqrt{3}}}}{\sqrt{3}}%
\phi^{\rho}\tilde{\partial}_{\rho}\tilde{\Phi}-e^{-\frac{\tilde{\Phi}}{\sqrt{%
3}}}\right)\right].  \label{eq:d(U-tilde)d(g-tilde)-SP}
\end{equation}

The effective energy-momentum tensor (\ref{eq:Tmunu-eff-metric-SP}) bears an
ordinary kinetic term for the scalar field plus terms coming from the
generalized potential $\tilde{U}$ which contains couplings between the
scalar and vector fields up to first order derivatives.


\subsection{Palatini formalism\label{subsec:SP-Palatini}}

In accordance with Eq.~(\ref{eq:S-Einstein-Pal}), Starobinsky-Podolsky
action in the Einstein frame and Palatini approach is:
\begin{equation}
S^{\prime\prime}=\int d^{4}x\sqrt{-\tilde{g}}\left[\tilde{R}-\tilde{U}\right] .
\label{eq:S-Einstein-Pal-SP}
\end{equation}
The potential $\tilde{U}$ in this equation assumes the form
\begin{equation}
\tilde{U}=\frac{1}{2c_{0}}\left(1+\frac{1}{\Phi}\tilde{\nabla}%
_{\mu}\phi^{\mu}-\frac{2}{\Phi^{2}}\phi^{\rho}\partial_{\rho}\Phi-\frac{1}{%
\Phi}\right)^{2}+\frac{1}{\Phi^{3}}\frac{\tilde{g}_{\alpha\beta}\phi^{%
\alpha}\phi^{\beta}}{2c_{1}},  \label{eq:U-tilde-Pal-SP}
\end{equation}
which is identical to the effective potential energy Eq.~(\ref%
{eq:U-tilde-metric-SP}) appearing in the metric formalism if one takes $%
\Phi=e^{\frac{\tilde{\Phi}}{\sqrt{3}}}$.

The EOM for $\Phi$ and $\phi^{\mu}$ are
\begin{equation}
\frac{1}{c_{0}}\left(\frac{2}{\Phi}\phi^{\rho}\tilde{\partial}_{\rho}\Phi-%
\tilde{\nabla}_{\rho}\phi^{\rho}-1\right)\left(\Phi+\tilde{\nabla}%
_{\sigma}\phi^{\sigma}-\frac{2}{\Phi}\phi^{\sigma}\tilde{\partial}%
_{\sigma}\Phi-1\right)-\frac{1}{\Phi}\frac{\tilde{g}_{\alpha\beta}\phi^{%
\alpha}\phi^{\beta}}{2c_{1}}=0  \label{eq:EOM-Pal-Phi-SP}
\end{equation}
and
\begin{equation}
\frac{1}{c_{0}}\tilde{\partial}_{\sigma}\left(\Phi+\tilde{\nabla}%
_{\rho}\phi^{\rho}-\frac{2}{\Phi}\phi^{\rho}\tilde{\partial}%
_{\rho}\Phi-1\right)-\frac{1}{\Phi}\frac{\phi_{\sigma}}{c_{1}}=0 .
\label{eq:EOM-Pal-(phi-mu)-SP}
\end{equation}
Even though the action Eq.~(\ref{eq:S-Einstein-Pal-SP}) does not contain a
canonical kinetic term for $\Phi$, the field equation Eq.~(\ref%
{eq:EOM-Pal-(phi-mu)-SP}) contains the second time derivative of $\Phi$ due
to the presence of the derivative coupling $\left(\frac{2}{\Phi^{2}}%
\phi^{\rho}\partial_{\rho}\Phi\right)^{2}$ in the potential energy. This
does not mean, however, $\Phi$ is a dynamical object. As one can see in Eq.~(%
\ref{eq:EOM-Pal-Phi-SP}), there is no second time derivative of any quantity
---{} it is a constraint equation. If an auxiliary vector field $%
V^{\rho}$ is introduced,
\begin{equation}
V^{\rho}\equiv\frac{\phi^{\rho}}{\Phi^{2}},  \label{eq:V}
\end{equation}
it is straightforward to show that
\begin{equation}
\tilde{\nabla}_{\rho}\phi^{\rho}-\frac{2}{\Phi}\phi^{\rho}\tilde{\partial}%
_{\rho}\Phi=\Phi^{2}\tilde{\nabla}_{\rho}V^{\rho}.  \label{eq:V_phi}
\end{equation}
Equations (\ref{eq:EOM-Pal-Phi-SP}) and (\ref{eq:EOM-Pal-(phi-mu)-SP}) can be
reexpressed as
\begin{eqnarray}
\Phi^{4}\left(\tilde{\nabla}_{\rho}V^{\rho}\right)^{2}+\Phi^{3}\left(\tilde{%
\nabla}_{\rho}V^{\rho}+\frac{c_{0}}{2c_{1}}\tilde{g}_{\alpha\beta}V^{%
\alpha}V^{\beta}\right)+\Phi-1 & =0  \label{eq:EOM-Pal-Phi-V}
\end{eqnarray}
and
\begin{equation}
\frac{1}{c_{0}}\frac{1}{\Phi}\tilde{\nabla}_{\sigma}\left(\Phi+\Phi^{2}%
\tilde{\nabla}_{\rho}V^{\rho}\right)-\frac{1}{c_{1}}V_{\sigma}=0,
\label{eq:EOM-Pal-V-SP}
\end{equation}
respectively. Equation (\ref{eq:EOM-Pal-Phi-V}) makes clear the nondynamical
character of the field $\Phi$ in the Palatini approach. Moreover, if one
replaces $\Phi$ derived from Eq.~(\ref{eq:EOM-Pal-Phi-V}) into Eq.~(\ref%
{eq:EOM-Pal-V-SP}), a nonlinear equation for $V^{\rho}$ is obtained. Its
analysis evidences that only the $V^{0}$ component is dynamical. An
analogous situation occurs in the Jordan frame, although the equations attained
in the latter case are distinct from the ones obtained here.

The gravitational field equation assumes, once more, the expected form of
Eq.~(\ref{eq:EOM-metric-g-SP}) but now
\begin{equation}
\tilde{T}_{\mu\nu}^{\left(\text{eff}\right)}=-\left(\tilde{g}_{\mu\nu}\tilde{%
U}-2\frac{\delta\tilde{U}}{\delta\tilde{g}^{\mu\nu}}\right)
\label{eq:Tmunu-eff-Pal-SP}
\end{equation}
with the second term calculated from Eq.~(\ref{eq:U-tilde-Pal-SP}) as
\begin{equation}
2\frac{\delta\tilde{U}}{\delta\tilde{g}^{\mu\nu}}=-\frac{1}{c_{1}\Phi^{3}}%
\phi_{\mu}\phi_{\nu}+\frac{1}{c_{0}}\tilde{g}_{\mu\nu}\tilde{\nabla}%
_{\lambda}\left[\frac{\phi^{\lambda}}{\Phi}\left(1+\frac{1}{\Phi}\tilde{%
\nabla}_{\rho}\phi^{\rho}-\frac{2}{\Phi^{2}}\phi^{\rho}\tilde{\partial}%
_{\rho}\Phi-\frac{1}{\Phi}\right)\right].
\label{eq:d(U-tilde)d(g-tilde)-Pal-SP}
\end{equation}

Now we turn to the study of the fluid representation for $\tilde{T}%
_{\mu\nu}^{\left(\text{eff}\right)}$ in both metric and Palatini formalisms.


\subsection{Fluid representation for $\tilde{T}_{\protect\mu\protect\nu%
}^{\left(\text{eff}\right)}$ \label{subsec:Fluid-Teff}}

A general energy-momentum tensor can be expressed in terms of an imperfect
fluid energy-momentum tensor \cite{Ellis2012}:
\begin{equation}
T_{\mu\nu}=\left(\varepsilon+p\right)u_{\mu}u_{\nu}+pg_{\mu\nu}+u_{\mu}q_{%
\nu}+u_{\nu}q_{\mu}+\pi_{\mu\nu},  \label{eq:Tmunu-fluid}
\end{equation}
where $p$ is the pressure, $\rho$ is the energy density, $u^{\mu}$ is the
four-velocity associated with a fluid element, $q_{\mu}$ denotes the heat flux,
and $\pi_{\mu\nu}$ stands for the viscous shear tensor. These various
quantities in Eq.~(\ref{eq:Tmunu-fluid}) satisfy the following properties:
\begin{equation}
\begin{cases}
q_{\mu}u^{\mu}=0, \\
\pi_{\mu\nu}u^{\nu}=0, \\
\pi_{\,\mu}^{\mu}=0, \\
\pi_{\mu\nu}=\pi_{\nu\mu},%
\end{cases}
\label{eq:Tmunu-fluid-properties}
\end{equation}
which determines the available degrees of freedom: two of them are related
to $\varepsilon$ and $p$, and three independent components in $T_{\mu\nu}$ come
from $q_{\mu}$ and five from $\pi_{\mu\nu}$. The metric determines the
background and the four-velocity fixes the reference system.

In this section, we shall encounter the form assumed by the energy-momentum
tensor $\tilde{T}_{\mu\nu}^{\left(\text{eff}\right)}$ for the
Starobinsky-Podolsky action. We begin by doing so in the metric formalism.


\subsubsection{$\tilde{T}_{\protect\mu\protect\nu}^{\left(\text{eff}\right)}$
in metric formalism\label{subsec:Teff-metric}}

In the metric formalism, the Starobinsky-Podolsky energy-momentum tensor is
shown to be of an imperfect fluid type with $\pi_{\mu\nu}=0$. This means
there are up to five independent components in $T_{\mu\nu}^{\left(\text{eff}%
\right)}$, namely one component related to $\Phi$ and four of them
associated with $\phi^{\mu}$.

In order to see that, we define the four-velocity as a linear combination of
the gradient of the scalar field and the vectorial field,
\begin{equation}
u_{\mu}=\frac{1}{N}\left(\partial_{\mu}\tilde{\Phi}+\chi\phi_{\mu}\right),%
\qquad N\equiv\sqrt{-\left(\partial_{\alpha}\tilde{\Phi}+\chi\phi_{\alpha}%
\right)\left(\partial^{\alpha}\tilde{\Phi}+\chi\phi^{\alpha}\right)},
\label{eq:4-velocity-metric}
\end{equation}
where $\chi$ is a quantity to be determined; $N$ is a normalization factor.

The next step is to decompose $\phi_{\mu}$ as
\begin{equation}
\phi_{\mu}=\lambda u_{\mu}+\phi^{\alpha}\left(\tilde{g}_{\alpha\mu}+u_{%
\alpha}u_{\mu}\right)=\lambda u_{\mu}+\frac{q_{\mu}}{\tau},
\label{eq:(phi-mu)(u,q)}
\end{equation}
which exhibits a component that is parallel to $u_{\mu}$ and a component
orthogonal to it. This means that the second term in Eq.~(\ref%
{eq:(phi-mu)(u,q)}) is parallel to the heat flux $q_{\mu}$ ---{} see
the first identity in Eq.~(\ref{eq:Tmunu-fluid-properties}). Therefore,
\begin{equation}
q_{\mu}\equiv\tau\phi^{\alpha}\left(\tilde{g}_{\alpha\mu}+u_{\alpha}u_{\mu}%
\right).  \label{eq:q-metric}
\end{equation}

We shall set
\begin{equation}
\chi=\frac{e^{-\frac{3\tilde{\Phi}}{2\sqrt{3}}}}{\sqrt{c_{1}}}
\label{eq:chi-metric}
\end{equation}
and
\begin{equation}
\tau=-N\chi.  \label{eq:tau-metric}
\end{equation}

The relations Eqs.~(\ref{eq:4-velocity-metric}), (\ref{eq:q-metric}) and
(\ref{eq:pi-metric}) can be substituted into Eq.~(\ref{eq:Tmunu-fluid}).
The resulting expression is then compared to Eq.~(\ref%
{eq:Tmunu-eff-metric-SP}), leading to the identifications
\begin{equation}
\left(\varepsilon+p\right)=N^{2}+2\chi\left(\partial_{\alpha}\tilde{\Phi}%
+\chi\phi_{\alpha}\right)\phi^{\alpha}  \label{eq:epsilon(p)-metric}
\end{equation}
and
\begin{equation}
p=-\frac{1}{2}\tilde{\partial}^{\alpha}\tilde{\Phi}\tilde{\partial}_{\alpha}%
\tilde{\Phi}-\tilde{U}+\frac{1}{c_{0}}\tilde{\nabla}_{\lambda}\left[%
\phi^{\lambda}e^{-\frac{\tilde{\Phi}}{\sqrt{3}}}\left(1+e^{-\frac{\tilde{\Phi%
}}{\sqrt{3}}}\tilde{\nabla}_{\rho}\phi^{\rho}-\frac{2e^{-\frac{\tilde{\Phi}}{%
\sqrt{3}}}}{\sqrt{3}}\phi^{\rho}\tilde{\partial}_{\rho}\tilde{\Phi}-e^{-%
\frac{\tilde{\Phi}}{\sqrt{3}}}\right)\right],  \label{eq:p-metric}
\end{equation}
with Eqs.~(\ref{eq:chi-metric}) and (\ref{eq:tau-metric}) completely
specifying the four-velocity Eq.~(\ref{eq:4-velocity-metric}) and the heat
flux Eq.~(\ref{eq:q-metric}).

Incidentally, the parameter $\lambda$ in Eq.~(\ref{eq:(phi-mu)(u,q)}) is
also found:
\begin{equation}
\lambda=-\frac{1}{N}\phi^{\alpha}\left(\partial_{\alpha}\tilde{\Phi}%
+\chi\phi_{\alpha}\right) . \label{eq:alpha-metric}
\end{equation}
Additionally, it becomes clear that there is no room for the viscous shear
tensor, i.e.
\begin{equation}
\pi_{\mu\nu}=0.  \label{eq:pi-metric}
\end{equation}
The potential $\text{$\tilde{U}$}$ shown up in the expression for $p$ is
given by Eq.~(\ref{eq:U-tilde-metric-SP}). By inserting Eq.~(\ref%
{eq:p-metric}) into Eq.~(\ref{eq:epsilon(p)-metric}), one obtains $%
\varepsilon$ in terms of the fields $\tilde{\Phi}$, $\phi^{\mu}$, and $\tilde{%
g}_{\mu\nu}$.

This reasoning completes our task of establishing, in the metric formalism,
Starobinsky-Podolsky $T_{\mu\nu}^{\left(\text{eff}\right)}$ as a shearless
imperfect fluid energy-momentum tensor.


\subsubsection{$\tilde{T}_{\protect\mu\protect\nu}^{\left(\text{eff}\right)}$
in Palatini formalism\label{subsec:Teff-Pal}}

The effective energy-momentum tensor in the Palatini formalism is given by Eqs.~(\ref{eq:Tmunu-eff-Pal-SP}), (\ref{eq:U-tilde-Pal-SP}) and (\ref{eq:d(U-tilde)d(g-tilde)-Pal-SP}), i.e.
\begin{equation}
\tilde{T}_{\mu\nu}^{\left(\text{eff}\right)}=-\frac{1}{c_{1}\Phi^{3}}%
\phi_{\mu}\phi_{\nu}+\tilde{g}_{\mu\nu}\left[-\tilde{U}+\frac{1}{c_{0}}%
\tilde{\nabla}_{\lambda}\left[\frac{\phi^{\lambda}}{\Phi}\left(1+\frac{1}{%
\Phi}\tilde{\nabla}_{\rho}\phi^{\rho}-\frac{2}{\Phi^{2}}\phi^{\rho}\tilde{%
\partial}_{\rho}\Phi-\frac{1}{\Phi}\right)\right]\right].
\label{eq:Tmunu-eff-Pal-SP-fluid}
\end{equation}
Direct comparison with Eq.~(\ref{eq:Tmunu-fluid}) sets it as a perfect fluid $%
T_{\mu\nu}$ with
\begin{equation}
q_{\mu}=\pi_{\mu\nu}=0.  \label{eq:q-pi-Pal}
\end{equation}
This comparison also yields
\begin{equation}
u_{\mu}=\frac{1}{N}\phi_{\mu},\qquad N\equiv\sqrt{-\phi_{\alpha}\phi^{\alpha}%
}  \label{eq:4-velocity-Pal}
\end{equation}
and
\begin{equation}
\left(\varepsilon+p\right)=-N^{2}\left(\frac{1}{c_{1}\Phi^{3}}\right),
\label{eq:epsilon(p)-Pal}
\end{equation}
with
\begin{equation}
p=-\tilde{U}+\frac{1}{c_{0}}\tilde{\nabla}_{\lambda}\left[\frac{%
\phi^{\lambda}}{\Phi}\left(1+\frac{1}{\Phi}\tilde{\nabla}_{\rho}\phi^{\rho}-%
\frac{2}{\Phi^{2}}\phi^{\rho}\tilde{\partial}_{\rho}\Phi-\frac{1}{\Phi}%
\right)\right] . \label{eq:p-Pal}
\end{equation}
Hence, Starobinsky-Podolsky $\tilde{T}_{\mu\nu}^{\left(\text{eff}\right)}$
is a perfect fluid tensor in the Palatini formalism.

The four-velocity Eq.~(\ref{eq:4-velocity-Pal}) is completely determined by
the auxiliary vector field $\phi^{\mu}$. From a geometrical point of view, $%
u_{\mu}$ scales as the gradient of the scalar curvature (given solely by $%
\phi^{\mu}$ in Palatini's case); this means the fluid velocity accompanies
the rate of change in the curvature. Moreover, since $u^{\mu}$ scales with $%
\phi^{\mu}$, this field defines the orientation of the $\tilde{T}%
_{\mu\nu}^{\left(\text{eff}\right)}$ decomposition. In particular, the $%
\phi^{0}$ component is the only one to produce a relevant contribution to
the direction of temporal evolution of the system under the choice of a
comoving frame $u^{\mu}=\left(1,0,0,0\right)$.

Equation (\ref{eq:epsilon(p)-Pal}) brings forth the possibility of violation of
the null energy condition \cite{Wald84} associated with what is known in
cosmology as the phantom regime, i.e. $p<-\varepsilon$. For $\Phi>0$, the
phantom regime takes place whenever the condition $c_{1}>0$ is satisfied. On
the other hand, this same condition leads to the presence of ghosts\footnote{%
A ghost is a field with kinetic energy with a ``wrong sign.''} in the
Starobinsky-Podolsky action \cite{Hindawi1996}. This demonstrates (at least
in this particular case) the direct relationship between ghosts and a
phantomlike behavior.


\section{Inflation in the Palatini formalism\label{sec:Inflation}}

Now an application to inflation is considered. The analysis is restrained to
the Palatini approach. The reason for this lies in the fact that the presence of
the imperfect fluid components demands a more careful analyzis in the metric
formalism ---{} this analysis is presented in \cite{StaPodInf}.

It is interesting to notice that the Starobinsky model in the Palatini approach
cannot generate an inflationary period, once the scalar auxiliary field is
not dynamical. Here the situation is different since the vector field could
eventually be responsible for driving inflation. This is addressed below.

Starobinsky-Podolsky action (\ref{eq:S-SP-geometric}) reduces to the
standard Starobinsky action in the limit $c_{1}\rightarrow 0$. The EOM (\ref%
{eq:EOM-Pal-Phi-SP}) and (\ref{eq:EOM-Pal-(phi-mu)-SP}) should be consistent
in this limit. In order to accomplish that we define a new vector field
\begin{equation}
\xi^{\mu}=\phi^{\mu}/c_{1}  \label{eq:xi}
\end{equation}
in terms of which the EOM read
\begin{align}
\frac{1}{c_{0}}\left(\frac{2}{\Phi}c_{1}\xi^{\rho}\partial_{\rho}\Phi-c_{1}%
\nabla_{\rho}\xi^{\rho}-1\right)\left(\Phi+c_{1}\nabla_{\sigma}\xi^{\sigma}-%
\frac{2}{\Phi}c_{1}\xi^{\sigma}\partial_{\sigma}\Phi-1\right)-\frac{c_{1}}{%
2\Phi}g_{\alpha\beta}\xi^{\alpha}\xi^{\beta} & =0,
\label{eq:EOM-Pal-Phi-SP-xi} \\
\frac{1}{c_{0}}\partial_{\sigma}\left(\Phi+c_{1}\nabla_{\rho}\xi^{\rho}-%
\frac{2}{\Phi}c_{1}\xi^{\rho}\partial_{\rho}\Phi-1\right)-\frac{1}{\Phi}%
g_{\sigma\rho}\xi^{\rho} & =0,  \label{eq:EOM-Pal-(phi-mu)-SP-xi}
\end{align}
where the tilde was omitted for notational economy.

Background cosmology is described by the Friedmann--Lema\^itre--Robertson--Walker (FLRW) line element,
\begin{equation}
ds^{2}=-dt^{2}+a^{2}\left( t\right) \left[ dr^{2}+r^{2}d\Omega ^{2}\right] ,
\label{eq:FLRW}
\end{equation}%
where $a\left( t\right) $ is the scale factor and $t$ is the coordinate time
(with $c=1$). The angular part is $d\Omega ^{2}=d\theta ^{2}+\sin ^{2}\theta
d\varphi ^{2}$. We consider a comoving reference frame: the four-velocity is
$u^{\mu }=\delta _{0}^{\mu }$. Homogeneity and isotropy of the space section
demand
\begin{equation}
\Phi =\Phi \left( t\right) \text{, } \quad \xi ^{\rho }=\left( \xi ^{0}\left(
t\right) ,0,0,0\right) \quad \Rightarrow \quad \nabla _{\mu }\xi ^{\mu }=\partial
_{0}\xi ^{0}+3H\xi ^{0} \quad \text{ and } \quad g_{\alpha \beta }\xi ^{\alpha }\xi
^{\beta }=-\left( \xi ^{0}\right) ^{2},  \label{eq:Phi(t)-xi0(t)}
\end{equation}%
with the usual definition
\begin{equation}
H\equiv \frac{1}{a}\frac{da}{dt}  \label{eq:H}
\end{equation}%
for the Hubble function. Then, Eqs.~(\ref{eq:EOM-Pal-Phi-SP-xi}) and
(\ref{eq:EOM-Pal-(phi-mu)-SP-xi}) turn to
\begin{align}
-\left[ c_{1}\left( \partial _{0}\xi ^{0}+3H\xi ^{0}-2\xi ^{0}\frac{\partial
_{0}\Phi }{\Phi }\right) +1\right] \left[ \Phi -1+c_{1}\left( \partial
_{0}\xi ^{0}+3H\xi ^{0}-2\xi ^{0}\frac{\partial _{0}\Phi }{\Phi }\right) %
\right] +\frac{c_{0}c_{1}}{2\Phi }\left( \xi ^{0}\right) ^{2}& =0,
\label{eq:EOM-Pal-Phi-SP-FRW} \\
\partial _{0}\Phi +c_{1}\partial _{0}\left( \partial _{0}\xi ^{0}+3H\xi
^{0}-2\xi ^{0}\frac{\partial _{0}\Phi }{\Phi }\right) +\frac{c_{0}}{\Phi }%
\xi ^{0}& =0.  \label{eq:EOM-Pal-(phi-mu)-SP-FRW}
\end{align}%
In terms of the new variable
\begin{equation}
Z=c_{1}\left( \partial _{0}\xi ^{0}+3H\xi ^{0}-2\xi ^{0}\frac{\partial
_{0}\Phi }{\Phi }\right) ,  \label{eq:Z}
\end{equation}%
the above equations are
\begin{align}
Z^{2}+\Phi Z+\left( \Phi -1\right) -\frac{c_{1}c_{0}}{2\Phi }\left( \xi
^{0}\right) ^{2}& =0,  \label{eq:EOM-Pal-Phi-SP-Z} \\
\partial _{0}\Phi +\partial _{0}Z+\frac{c_{0}}{\Phi }\xi ^{0}& =0.
\label{eq:EOM-Pal-(phi-mu)-SP-Z}
\end{align}%
In the limit $c_{1}\rightarrow 0$, one has $Z\rightarrow 0$ and $\Phi
\rightarrow 1$.

The next step is to write down the Friedmann equations (for the perfect fluid),
\begin{align}
H^{2}& =\frac{1}{3M_{Pl}^{2}}\varepsilon =\frac{1}{6}\varepsilon ,
\label{eq:Fried-H2} \\
\frac{dH}{dt}& =-\frac{1}{2M_{Pl}^{2}}\left( \varepsilon +p\right) =-\frac{1%
}{4}\left( \varepsilon +p\right) ,  \label{eq:Fried-Hdot}
\end{align}%
where $16\pi G=1\Rightarrow M_{Pl}^{2}=2$. Taking into account Eqs.~(\ref%
{eq:U-tilde-Pal-SP}), (\ref{eq:epsilon(p)-Pal}), (\ref{eq:p-Pal}) and using
the definition (\ref{eq:xi}), the quantities $\varepsilon $ and $p$ are
written as
\begin{equation}
\varepsilon +p=c_{1}\frac{g_{\alpha \beta }\xi ^{\beta }\xi ^{\alpha }}{\Phi
^{3}}  \label{eq:epsilon(p)-Pal-xi}
\end{equation}%
and
\begin{align}
p=& -U+\frac{c_{1}}{c_{0}}\left( \Phi +c_{1}\nabla _{\rho }\xi ^{\rho }-%
\frac{2c_{1}}{\Phi }\xi ^{\rho }\partial _{\rho }\Phi -1\right) \left( \frac{%
1}{\Phi ^{2}}\nabla _{\lambda }\xi ^{\lambda }-2\frac{\xi ^{\lambda }}{\Phi
^{2}}\frac{\partial _{\lambda }\Phi }{\Phi }\right)  \notag \\
& +\frac{c_{1}}{c_{0}}\frac{\xi ^{\lambda }}{\Phi ^{2}}\partial _{\lambda
}\left( \Phi +c_{1}\nabla _{\rho }\xi ^{\rho }-\frac{2c_{1}}{\Phi }\xi
^{\rho }\partial _{\rho }\Phi -1\right) ,  \label{eq:p-Pal-xi}
\end{align}%
where
\begin{equation}
U=\frac{1}{2c_{0}\Phi ^{2}}\left( \Phi +c_{1}\nabla _{\mu }\xi ^{\mu
}-2c_{1}\xi ^{\rho }\frac{\partial _{\rho }\Phi }{\Phi }-1\right) ^{2}+\frac{%
c_{1}}{2\Phi ^{3}}g_{\alpha \beta }\xi ^{\alpha }\xi ^{\beta }.
\label{eq:U-xi}
\end{equation}%
Inserting Eq.~(\ref{eq:EOM-Pal-(phi-mu)-SP-FRW}) into the last term of Eq.~(\ref%
{eq:p-Pal-xi})
\begin{equation}
p=-U+\frac{1}{c_{0}\Phi ^{2}}\left( c_{1}\nabla _{\rho }\xi ^{\rho
}-2c_{1}\xi ^{\rho }\frac{\partial _{\rho }\Phi }{\Phi }\right) ^{2}+\frac{1%
}{c_{0}\Phi ^{2}}\left( c_{1}\nabla _{\lambda }\xi ^{\lambda }-2c_{1}\xi
^{\lambda }\frac{\partial _{\lambda }\Phi }{\Phi }\right) \left( \Phi
-1\right) +\frac{c_{1}}{\Phi ^{3}}g_{\sigma \rho }\xi ^{\rho }\xi ^{\sigma };
\label{eq:p-Pal-xi-EOM}
\end{equation}%
so that Eq.~(\ref{eq:epsilon(p)-Pal-xi}) turns to
\begin{equation}
\varepsilon =\frac{1}{2c_{0}\Phi ^{2}}\left( \Phi -1\right) ^{2}-\frac{%
c_{1}^{2}}{2c_{0}\Phi ^{2}}\left( \nabla _{\mu }\xi ^{\mu }-2\xi ^{\rho }%
\frac{\partial _{\rho }\Phi }{\Phi }\right) ^{2}+\frac{c_{1}}{2\Phi ^{3}}%
g_{\alpha \beta }\xi ^{\alpha }\xi ^{\beta }.  \label{eq:epsilon-Pal-xi}
\end{equation}%
Particularly, on the FLRW background, Eqs.~(\ref{eq:epsilon(p)-Pal-xi}) and (\ref%
{eq:epsilon-Pal-xi}) are
\begin{equation}
\varepsilon +p=-c_{1}\frac{\left( \xi ^{0}\right) ^{2}}{\Phi ^{3}}
\label{eq:epsilon(p)-Pal-FRW}
\end{equation}%
and
\begin{equation}
\varepsilon =\frac{1}{2c_{0}\Phi ^{2}}\left( \Phi -1\right) ^{2}-\frac{1}{%
2c_{0}c_{1}\Phi ^{2}}Z^{2}-\frac{c_{1}}{2\Phi ^{3}}\left( \xi ^{0}\right)
^{2}.  \label{eq:epsilon-Pal-FRW}
\end{equation}%
Therefore, Friedmann equations (\ref{eq:Fried-H2}) and (\ref{eq:Fried-Hdot})
are cast into the form
\begin{align}
H^{2}& =\frac{1}{6}\left[ \frac{1}{2c_{0}\Phi ^{2}}\left( \Phi -1\right)
^{2}-\frac{1}{2c_{0}\Phi ^{2}}Z^{2}-\frac{c_{1}}{2\Phi ^{3}}\left( \xi
^{0}\right) ^{2}\right] ,  \label{eq:Fried-H2-xi} \\
\frac{dH}{dt}& =\frac{c_{1}}{4}\frac{\left( \xi ^{0}\right) ^{2}}{\Phi ^{3}}.
\label{eq:fried-Hdot-xi}
\end{align}%
Background cosmology in the context of our model is described by Eqs.~(\ref%
{eq:EOM-Pal-Phi-SP-Z}), (\ref{eq:EOM-Pal-(phi-mu)-SP-Z}), (\ref%
{eq:Fried-H2-xi}) and (\ref{eq:fried-Hdot-xi}).

It is convenient to write the various equations in terms of dimensionless
quantities. By a simple inspection of the action, we are able to associate
each quantity with its dimension: $c_{0}\rightarrow \left[ \text{mass}\right]
{}^{-2}$; $c_{1}\rightarrow \left[ \text{mass}\right] ^{-4}$; $\phi ^{\mu
}\rightarrow \left[ \text{mass}\right] ^{-1}$; $\Phi \rightarrow \left[
\text{mass}\right] ^{0}$; $\xi ^{\mu }\rightarrow \left[ \text{mass}\right]
^{3}$. Accordingly, let us define
\begin{align}
h_{c}& \equiv c_{0}^{1/2}H\Rightarrow H=c_{0}^{-1/2}h_{c} ,  \label{eq:hc} \\
\Theta & \equiv c_{0}^{3/2}\xi ^{0}\Rightarrow \xi ^{0}=\Theta c_{0}^{-3/2} ,
\label{eq:Theta} \\
\dot{Q}& \equiv \bar{\partial}_{0}Q=c_{0}^{1/2}\partial _{0}Q ,
\label{eq:Qdot} \\
\beta & \equiv -\frac{c_{1}}{c_{0}^{2}} ,  \label{eq:beta}
\end{align}%
in terms of which Eqs.~(\ref{eq:EOM-Pal-Phi-SP-Z}), (\ref%
{eq:EOM-Pal-(phi-mu)-SP-Z}), (\ref{eq:Fried-H2-xi}), and (\ref%
{eq:fried-Hdot-xi}) become
\begin{align}
h_{c}^{2}& =\frac{1}{6}\left[ \frac{\beta }{2\Phi ^{3}}\Theta ^{2}+\frac{1}{2%
}\left( 1-\frac{1}{\Phi }\right) ^{2}-\frac{1}{2\Phi ^{2}}Z^{2}\right]
\label{FRW adm 1a} , \\
\dot{h}_{c}& =-\frac{\beta }{4}\frac{\Theta ^{2}}{\Phi ^{3}}
\label{FRW adm 2a}
\end{align}
and
\begin{align}
Z^{2}+\Phi Z+\left( \Phi -1\right) +\frac{\beta }{2\Phi }\Theta ^{2}& =0,
\label{Field adm 1a} \\
\dot{Z}+\dot{\Phi}+\frac{\Theta }{\Phi }& =0,  \label{Field adm 2a}
\end{align}%
where [cf. Eq.~(\ref{eq:Z})]
\begin{equation}
Z=-\beta \left( \dot{\Theta}+3h_{c}\Theta -2\Theta \frac{\dot{\Phi}}{\Phi }%
\right) .  \label{Def Z}
\end{equation}%
Obviously, $\beta =0\Rightarrow Z=0$; then, Eq.~(\ref{Field adm 1a}) imposes
$\Phi =1$. (This was checked before from the limit $c_{1}\rightarrow 0$.) In
this case, we get $h_{c}=0$ implying a constant scale factor. Hence, the
standard Starobinsky action does not engender dynamics in the Palatini
formalism.

The next step is to use Eq.~(\ref{Field adm 1a}) to write $Z$ as
\begin{equation}
2Z=-\Phi +\sqrt{\left( \Phi -2\right) ^{2}-\frac{2\beta }{\Phi }\Theta ^{2}} ,
\label{Sol de Z}
\end{equation}%
where the solution with a negative sign multiplying the square root had to
be neglected. This is needed if we claim the solution $\Phi =1$ to be recovered in the limit
$\beta \rightarrow 0$. By differentiating this expression, replacing the result
into Eq.~(\ref{Field adm 2a}), and performing some manipulations, we
obtain the final set of equations which will be used to analyze the
inflationary solutions:
\begin{align}
h_{c}^{2}\left(\Phi,\Theta\right) & =\frac{1}{6}\left[\frac{\beta}{2}\frac{%
\Theta^{2}}{\Phi^{3}}+\frac{1}{2}\left(1-\frac{1}{\Phi}\right)^{2}-\frac{%
\left(J-\Phi\right)^{2}}{8\Phi^{2}}\right] ,  \label{h Phi e Theta} \\
\dot{h}_{c}\left(\Phi,\Theta\right) & =-\frac{\beta}{4}\frac{\Theta^{2}}{%
\Phi^{3}}  \label{h pto Phi e Theta}
\end{align}
and
\begin{align}
\dot{\Phi} & \equiv f\left(\Theta,\Phi\right)=\frac{\Theta}{\Phi}\left(\frac{%
-3J+\Phi-6\beta h_{c}\Theta}{J+\left(\Phi-2\right)-3\beta\left(\frac{\Theta}{%
\Phi}\right)^{2}}\right),  \label{dPhi dt} \\
\dot{\Theta} & \equiv g\left(\Theta,\Phi\right)=\frac{\Phi-J}{2\beta}%
-3h_{c}\Theta+2\frac{\Theta^{2}}{\Phi^{2}}\left(\frac{-3J+\Phi-6\beta
h_{c}\Theta}{J+\left(\Phi-2\right)-3\beta\left(\frac{\Theta}{\Phi}\right)^{2}%
}\right),  \label{dTheta dt}
\end{align}
where
\begin{equation}
J\left(\Phi,\Theta\right)=\Phi\sqrt{\left(1-\frac{2}{\Phi}\right)^{2}-2\frac{%
\Theta^{2}}{\Phi^{3}}\beta}.  \label{J Phi e Theta}
\end{equation}

From Eq.~(\ref{h pto Phi e Theta}) we see that
\begin{equation}
\frac{\ddot{a}}{a}=-\frac{\beta }{4\Phi ^{3}}\Theta ^{2}+h_{c}^{2}\text{.}
\label{a2dots}
\end{equation}%
Hence, a decelerated regime for the scale factor (necessary for the end inflation)
can be obtained only if $\beta $ and $\Phi $ have the
same sign (i.e. $\Phi \beta \geq 0$), once $h_{c}^{2}$ is always positive.
In principle, a transition from a positive (negative) to a negative
(positive) $\Phi $ could be considered. However, when crossing $\Phi =0$
divergences show up, for instance, in $\dot{h}_{c}$ and $h_{c}^{2}$. This
is the reason why $\Phi $ will be restricted to have the same sign as $\beta $. Note that
this condition together with the fact that $J\left( \Phi ,\Theta \right) $
has to be a real quantity implies further restrictions on the parameters
$\left\{ \Phi _{i},\Theta _{i}\right\} $. In fact, a real $J$ is obtained only if
\begin{equation}
\left( 1-\frac{2}{\Phi }\right) ^{2}\geq 2\frac{\Theta ^{2}}{\Phi ^{3}}\beta
.  \label{rest. esp. de fase}
\end{equation}%
We emphasize that this condition must be satisfied throughout the
evolution of the inflationary period. Also we realize that $\Phi =2$ is only
attainable if $\Theta =0$, otherwise $J$ would be complex.


\subsection{Inflationary regime}

In this section, we analyze which values of the parameters $\left\{
\Phi_{i},\Theta_{i},\beta\right\} $ (where the index $i$ stands for initial
conditions) could give rise to a quasiexponential expansion followed by a
radiationlike decelerated universe (hot big bang scenario).

The conditions (\ref{rest. esp. de fase}) and $\Phi\beta\ge0$ essentially
establish that the quantity $\epsilon_{H}$, as defined below, is a real
(positive) quantity:
\begin{equation}
\epsilon_{H}\equiv\frac{-\dot{h}_{c}}{h_{c}^{2}}>0\text{.}
\label{Slow-roll 1}
\end{equation}
This parameter is intimately related to the existence of an inflationary
regime. In order to obtain a quasiexponential expansion, a necessary
condition is that $\epsilon_{H}\ll1$. Due to Eqs.~(\ref{h Phi e Theta})
and (\ref{h pto Phi e Theta}), this condition is obtained only if
\begin{equation}
\left(1-\frac{1}{\Phi}\right)^{2}\gg\beta\frac{\Theta^{2}}{\Phi^{3}}.
\label{Cond de Inflacao}
\end{equation}
If we assume (\ref{Cond de Inflacao}) is satisfied and exclude the possibility of
$\Phi $ taking values in the neighborhood of $\Phi =2$,\footnote{In general, $\Phi \approx 2$ does not satisfy (\ref{rest. esp. de fase}).} then Eq.~(\ref{J Phi e Theta}) can be approximated by
\begin{equation}
J\simeq \Phi \left\vert 1-\frac{2}{\Phi }\right\vert \left( 1-\frac{\frac{%
\Theta ^{2}}{\Phi ^{3}}\beta }{\left( 1-\frac{2}{\Phi }\right) ^{2}}\right) .
\label{Japprox}
\end{equation}%
As a consequence,
\begin{equation}
\epsilon _{H}\simeq \frac{\frac{\beta }{4}\frac{\Theta ^{2}}{\Phi ^{3}}}{%
\frac{1}{6}\left[ \frac{\beta }{2}\frac{\Theta ^{2}}{\Phi ^{3}}+\frac{1}{2}%
\left( 1-\frac{1}{\Phi }\right) ^{2}-\frac{1}{8}\left( \left\vert 1-\frac{2}{%
\Phi }\right\vert \left( 1-\frac{\frac{\Theta ^{2}}{\Phi ^{3}}\beta }{\left(
1-\frac{2}{\Phi }\right) ^{2}}\right) -1\right) ^{2}\right] }.
\label{epsapprox0}
\end{equation}

We separate the analyzis in two cases:
\begin{enumerate}
\item[i)] $0<\Phi <2$. In this case,
\begin{equation*}
\epsilon _{H}\simeq \frac{3}{\left[ 1+\frac{\left( 1-\frac{1}{\Phi }\right)
}{\left( 1-\frac{2}{\Phi }\right) }\right] }.
\end{equation*}%
The condition $\epsilon _{H}\ll 1$ is satisfied only if $\Phi \approx 2$.
However, this is not in agreement with the assumption made above.
We conclude that no inflation can take place in this case.

\item[ii)] $\Phi <0$ or $\Phi >2$. We find
\begin{equation*}
\epsilon _{H}\simeq \frac{\frac{\beta }{4}\frac{\Theta ^{2}}{\Phi ^{3}}}{%
\frac{1}{6}\left[ \frac{\beta }{2}\frac{\Theta ^{2}}{\Phi ^{3}}+\frac{1}{2}%
\left( 1-\frac{1}{\Phi }\right) ^{2}-\frac{1}{2\Phi ^{2}}-\frac{1}{2\Phi }%
\frac{\frac{\Theta ^{2}}{\Phi ^{3}}\beta }{\left( 1-\frac{2}{\Phi }\right) }%
\right] }.
\end{equation*}%
If we exclude the neighborhoods of $\Phi =0$ and $\Phi =2$, we can
approximate the expression above as
\begin{equation}
\epsilon _{H}\simeq \frac{3\beta \Theta ^{2}}{\Phi ^{3}\left( 1-\frac{2}{%
\Phi }\right) }.  \label{epsapprox3}
\end{equation}
\end{enumerate}

For an inflationary regime, the condition $\epsilon _{H}\simeq \frac{3\beta
\Theta ^{2}}{\Phi ^{2}\left( \Phi -2\right) }\ll 1$ must be satisfied
simultaneously with Eq.~(\ref{Cond de Inflacao}). This is achieved if
\begin{equation}
\frac{\Theta }{\Phi }\ll 1\Rightarrow \frac{\beta \Theta ^{2}}{\Phi ^{3}}\ll
1,  \label{Cond de Inflacao 1}
\end{equation}%
as long as $\frac{\beta }{\Phi }$ is not so large. The next step is to
check if the conditions used so far suggest the existence of an
inflationary scenario. In order to do so, we study the second condition for
inflation, which reads
\begin{equation}
\eta _{H}=-\frac{\dot{\epsilon}_{H}}{h_{c}\epsilon _{H}}\ll 1.
\label{Slow-roll 2}
\end{equation}%
Equations (\ref{dPhi dt}) and (\ref{dTheta dt}) can be rewritten under condition (\ref{Cond de Inflacao 1}) enabling us to calculate $\dot{\epsilon_H}$. Since $J$ and $h_{c}^{2}$ become, respectively, $J\simeq \Phi
-2$ and $h_{c}^{2}\simeq \frac{1}{12}\left( 1-\frac{2}{\Phi }\right) $, we
end up with
\begin{align}
\dot{\Phi}& \simeq -\frac{\Theta }{\Phi }\left( \frac{1-\frac{3}{\Phi }}{1-%
\frac{2}{\Phi }}\right) ,  \label{dPhi dt aprox} \\
\dot{\Theta}& \simeq \frac{1}{\beta }-\sqrt{\frac{3}{4}\left( 1-\frac{2}{%
\Phi }\right) }\Theta ,  \label{dTheta dt aprox}
\end{align}%
and
\begin{equation}
\eta _{H}\simeq -\frac{2\sqrt{3}}{\sqrt{1-\frac{2}{\Phi }}}\left[ \frac{1}{%
\Theta }\left( \frac{2}{\beta }-\sqrt{3\left( 1-\frac{2}{\Phi }\right) }%
\Theta \right) +4\frac{\Theta }{\Phi ^{2}}\left( \frac{1-\frac{3}{\Phi }}{1-%
\frac{2}{\Phi }}\right) \left[ 1+\frac{3\left( 1-\frac{4}{3\Phi }\right) }{%
4\left( 1-\frac{2}{\Phi }\right) }\right] \right] . \label{eq:etaH}
\end{equation}

From Eqs.~(\ref{dPhi dt aprox}) and (\ref{Cond de Inflacao 1}) we conclude
that $\left|\dot{\Phi}\right|\ll1$. This means $\Phi$ varies slowly and can
be taken approximately as a constant in Eq.~(\ref{dTheta dt aprox}). As a
consequence
\begin{equation}
\Theta\left(t\right)\simeq e^{-\frac{t}{2}\sqrt{3\left(1-\frac{2}{\Phi}%
\right)}}+\frac{2}{\beta\sqrt{3\left(1-\frac{2}{\Phi}\right)}}.
\label{Thetat}
\end{equation}
The exponential decay shows that after some short time
\begin{equation}
\Theta\simeq\frac{2}{\beta\sqrt{3\left(1-\frac{2}{\Phi}\right)}},
\label{Thetainf}
\end{equation}
which is an accumulation point on the $(\Theta,\dot{\Theta})$ plane. This reduces Eq.~(\ref{eq:etaH}) to
\begin{equation}
\eta_{H}\simeq-\frac{\Theta}{\Phi}\frac{8\sqrt{3}}{\sqrt{1-\frac{2}{\Phi}}}%
\frac{1}{\Phi}\left(\frac{1-\frac{3}{\Phi}}{1-\frac{2}{\Phi}}\right)\left[1+%
\frac{3\left(1-\frac{4}{3\Phi}\right)}{4\left(1-\frac{2}{\Phi}\right)}\right]%
\ll1.  \label{etaapprox0}
\end{equation}

Up to this point, we concluded that the conditions $\left( \frac{\Theta }{\Phi }\ll 1 \right)$
and $( \Phi <0 \,\,\, \text{or} \,\,\, \Phi >2 )$ imply $\epsilon _{H}\ll 1$ and $\eta _{H}\ll 1$,
which means inflation occurs. The question that arises now is whether
this inflation ends. In order to answer this question, we manipulate Eqs.~(\ref{dPhi
dt aprox}) and (\ref{dTheta dt aprox}) to obtain the expression that allows
us to plot the direction fields on the $(\Phi, \dot{\Phi})$ phase space:
\begin{equation}
\frac{d\dot{\Phi}}{d\Phi }\simeq -\frac{1}{\beta }\frac{1}{\dot{\Phi}\Phi }%
\left( \frac{1-\frac{3}{\Phi }}{1-\frac{2}{\Phi }}\right) -\sqrt{\frac{3}{4}%
\left( 1-\frac{2}{\Phi }\right) }+\frac{\dot{\Phi}}{\Phi }+\dot{\Phi}\frac{3%
}{\Phi ^{2}}\frac{1}{\left( 1-\frac{3}{\Phi }\right) }-\frac{2}{\Phi
^{2}\left( 1-\frac{2}{\Phi }\right) }\dot{\Phi}.  \label{DirecfieldPhi}
\end{equation}

Now let us consider the case $\left\vert \Phi \right\vert \gg 1$. Equations (\ref%
{dPhi dt aprox}), (\ref{dTheta dt aprox}), and (\ref{DirecfieldPhi}) are reduced to
\begin{equation}
\begin{cases}
\dot{\Phi}\simeq -\frac{\Theta }{\Phi }, \\
\dot{\Theta}\simeq \frac{1}{\beta }-\sqrt{\frac{3}{4}}\Theta, %
\end{cases}
\label{PhidotThetadotApprox}
\end{equation}%
and
\begin{equation}
\frac{d\dot{\Phi}}{d\Phi }\simeq -\frac{1}{\Phi \dot{\Phi}\beta }-\frac{\dot{%
\Phi}}{\Phi }-\sqrt{\frac{3}{4}}. \label{ddotPhi/dPhi}
\end{equation}%
With Eqs.~(\ref{PhidotThetadotApprox}) and (\ref{ddotPhi/dPhi}) we can easily analyze the direction fields for both $\Phi _{i}<0$ and $%
\Phi _{i}>2$ cases:
\begin{enumerate}
\item[i)] $\Phi _{i}<0$. Figure \ref{fig:DFPhisNegative} shows the plot for the direction
fields on the $(\Phi, \dot{\Phi})$ phase space for $\Phi
_{i}<0$. We note the existence of an attractor line which indicates the
slow-roll regime for the field $\Phi $. This line, however, shows $\Phi $
decreasing while $\dot{\Phi}$ tends to zero. This behavior indicates that
although inflation takes place, it apparently does not end, since the ratio $%
\frac{\Theta }{\Phi }$ becomes even smaller.
This statement can be verified when Eq.~(\ref{Thetainf}) is backsubstituted into
(\ref{dPhi dt aprox}). We obtain
\begin{equation}
\dot{\Phi}\simeq -\frac{2}{\Phi \beta \sqrt{3\left( 1-\frac{2}{\Phi }\right)
}}\left( \frac{1-\frac{3}{\Phi }}{1-\frac{2}{\Phi }}\right) <0\text{ with }%
\left\vert \dot{\Phi}\right\vert \ll 1.  \label{Phipontpapprox}
\end{equation}%
Since $\beta $ and $\Phi $ have the same sign, then $\dot{\Phi}<0$. This
way, if the initial condition is such that $\Phi _{i}<0-\Delta $, with $%
\Delta $ being positive (which means $\beta $ is negative), then the
condition $\frac{\Theta }{\Phi }\ll 1$ sets $\Theta $ to be negative. When $%
\Theta $ reaches the accumulation point, $\Phi $ decreases (slowly),
increasing its modulus. This is such that the ratio $\frac{\Theta }{\Phi }$ becomes
smaller and smaller. Then the conditions for inflation never cease and
inflation cannot end.

\begin{figure}[ht]
\par
(a)\includegraphics[scale=0.4]{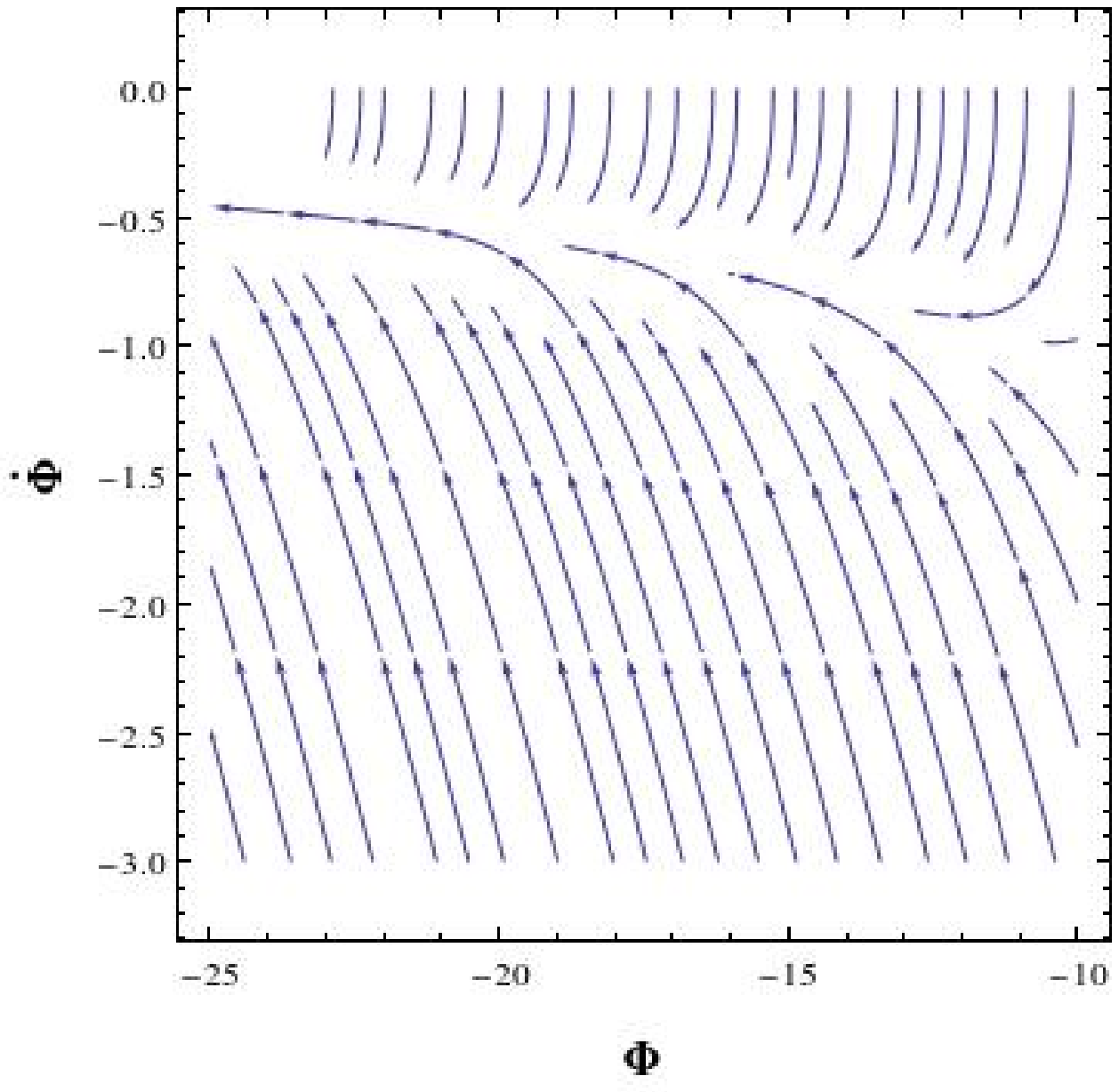} ~%
(b)\includegraphics[scale=0.4]{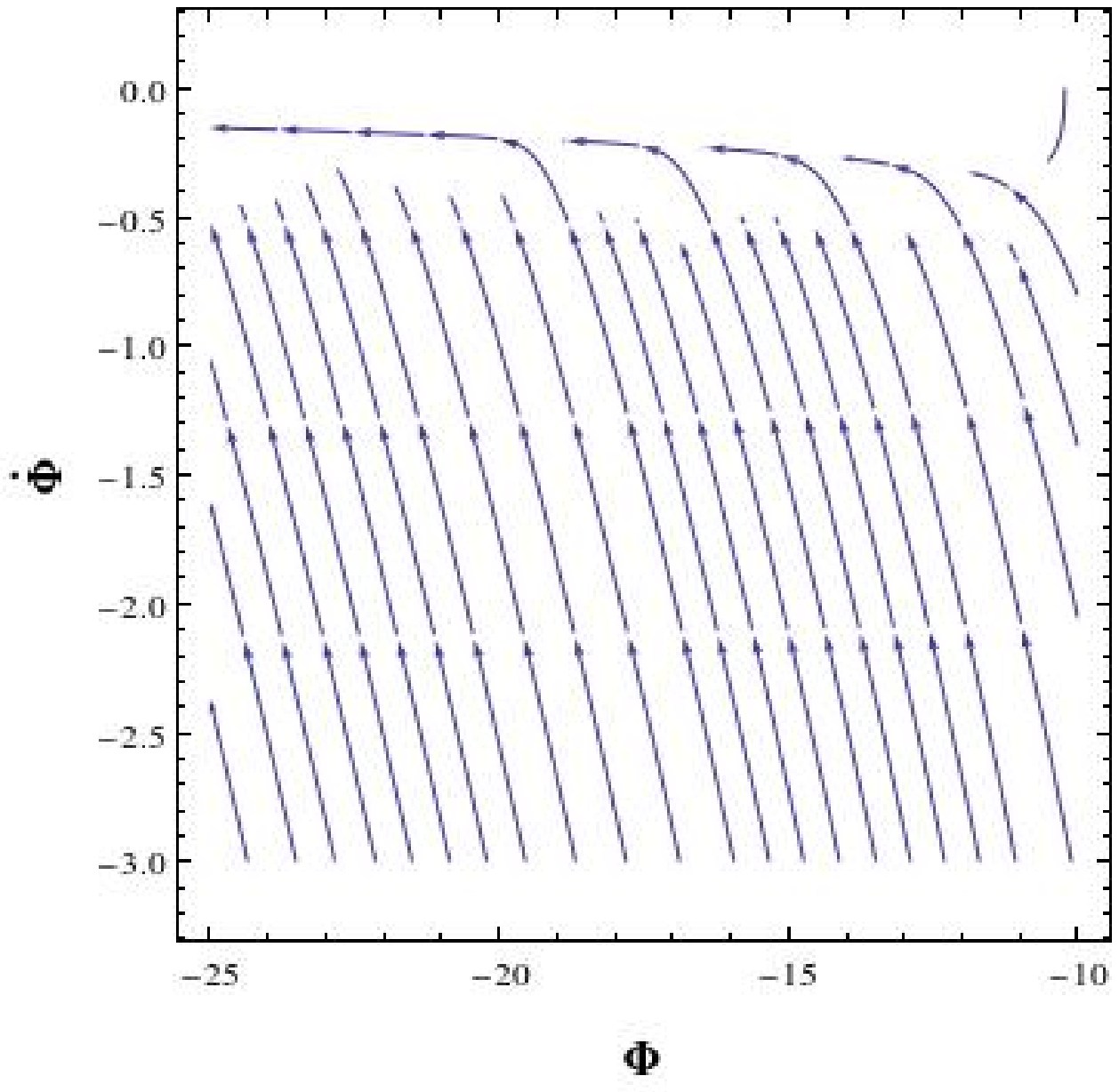}~~%
(c)\includegraphics[scale=0.4]{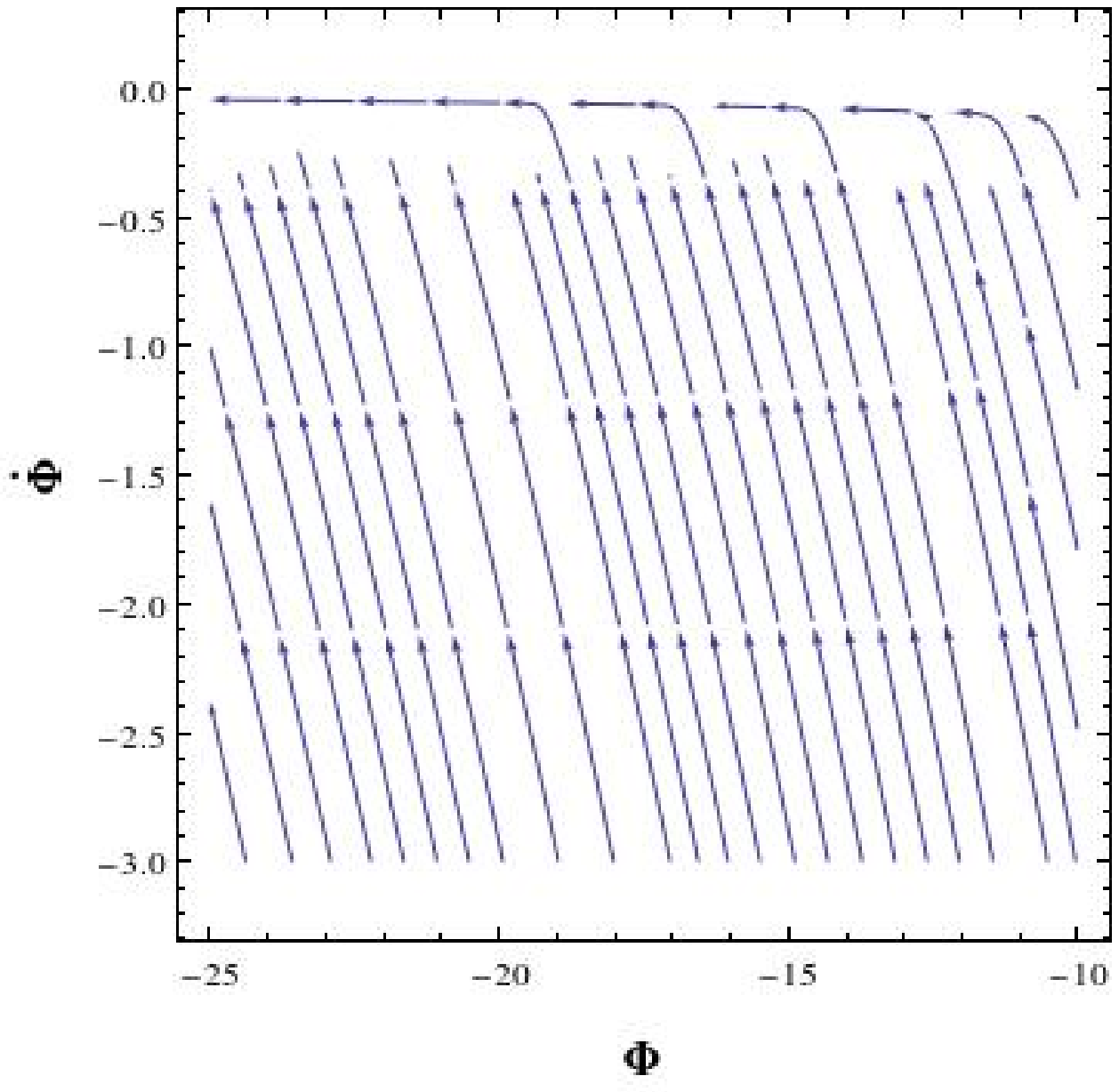}
\caption{Direction fields on the $(\Phi, \dot{\Phi})$ phase space for
negative values of $\Phi $ and various values of $\protect\beta $: (a) $%
\protect\beta =-0.1$, (b) $\protect\beta =-0.3$, and (c) $\protect\beta =-1$.}
\label{fig:DFPhisNegative}
\end{figure}

\item[ii)] $\Phi _{i}>2$. The direction fields on this approximative
analysis are shown in Fig.~\ref{fig:DFPhisPositive}. We notice the existence
of an attractor line for different values of $\beta $. This indicates the
existence of the slow-roll regime. The attractor line shows $\Phi $
decreasing with time. If we extrapolate the observed behavior of this line,
we expect the field $\Phi $ to reach values for which the condition $\frac{%
\Theta }{\Phi }\ll 1$ will eventually break down. This is a necessary condition
for inflation to cease. However, the condition $\Phi \gg 1$ will also be
violated and when this happens, we cannot rely on Eqs.~(\ref%
{PhidotThetadotApprox}) to properly describe our system. 

\begin{figure}[ht]
\par
(a)\includegraphics[scale=0.4]{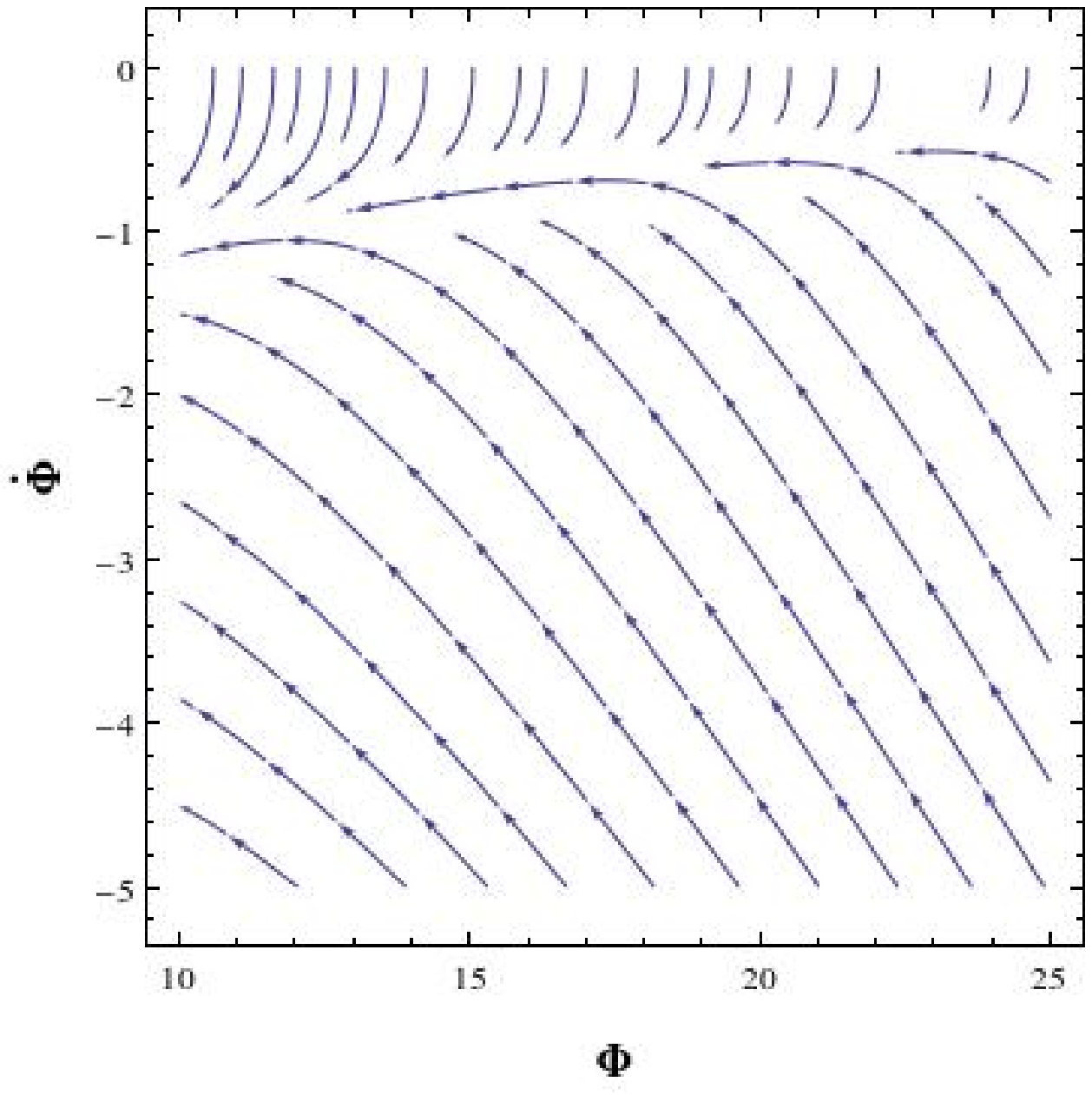} ~%
(b)\includegraphics[scale=0.4]{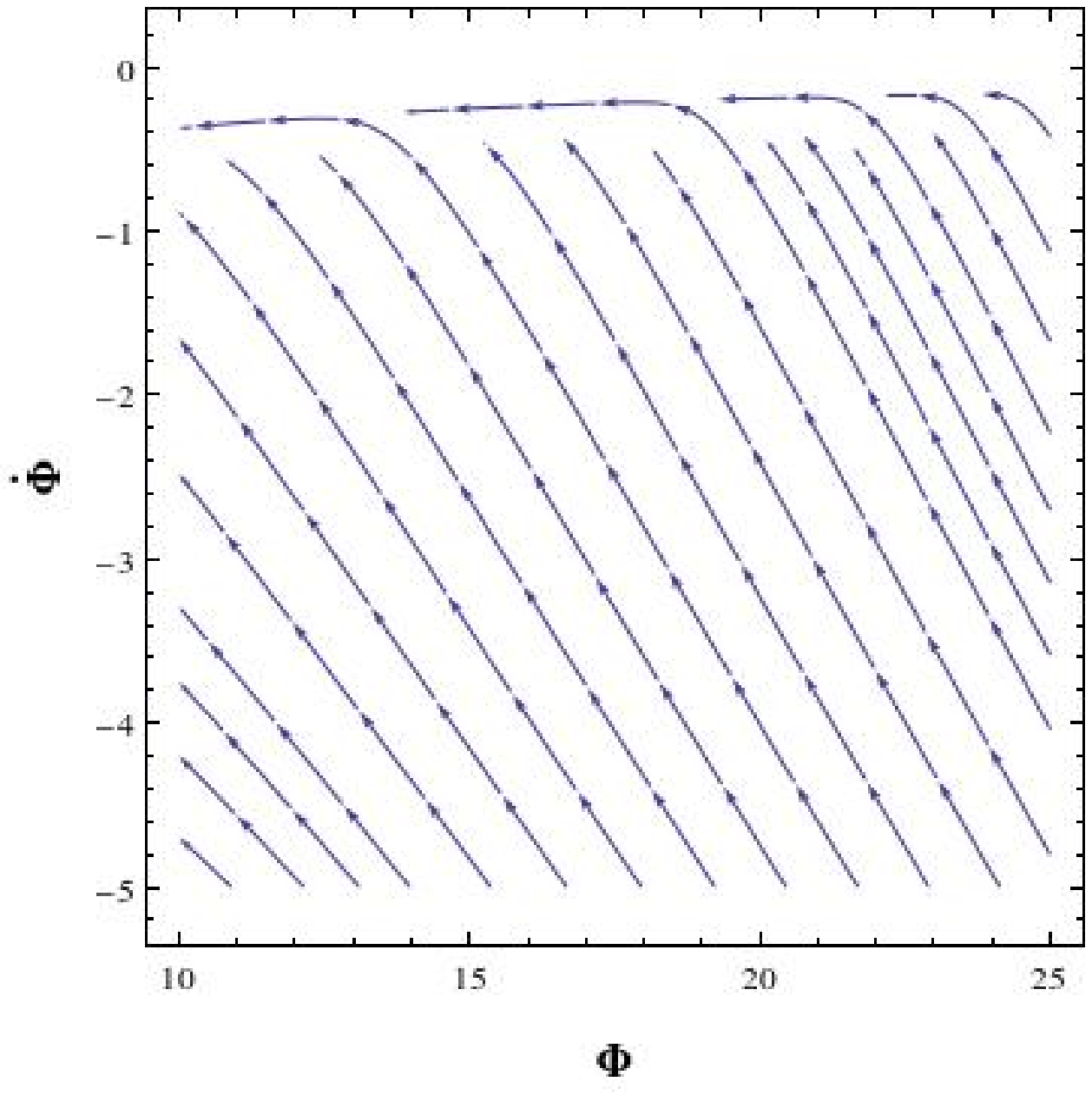}~~%
(c)\includegraphics[scale=0.4]{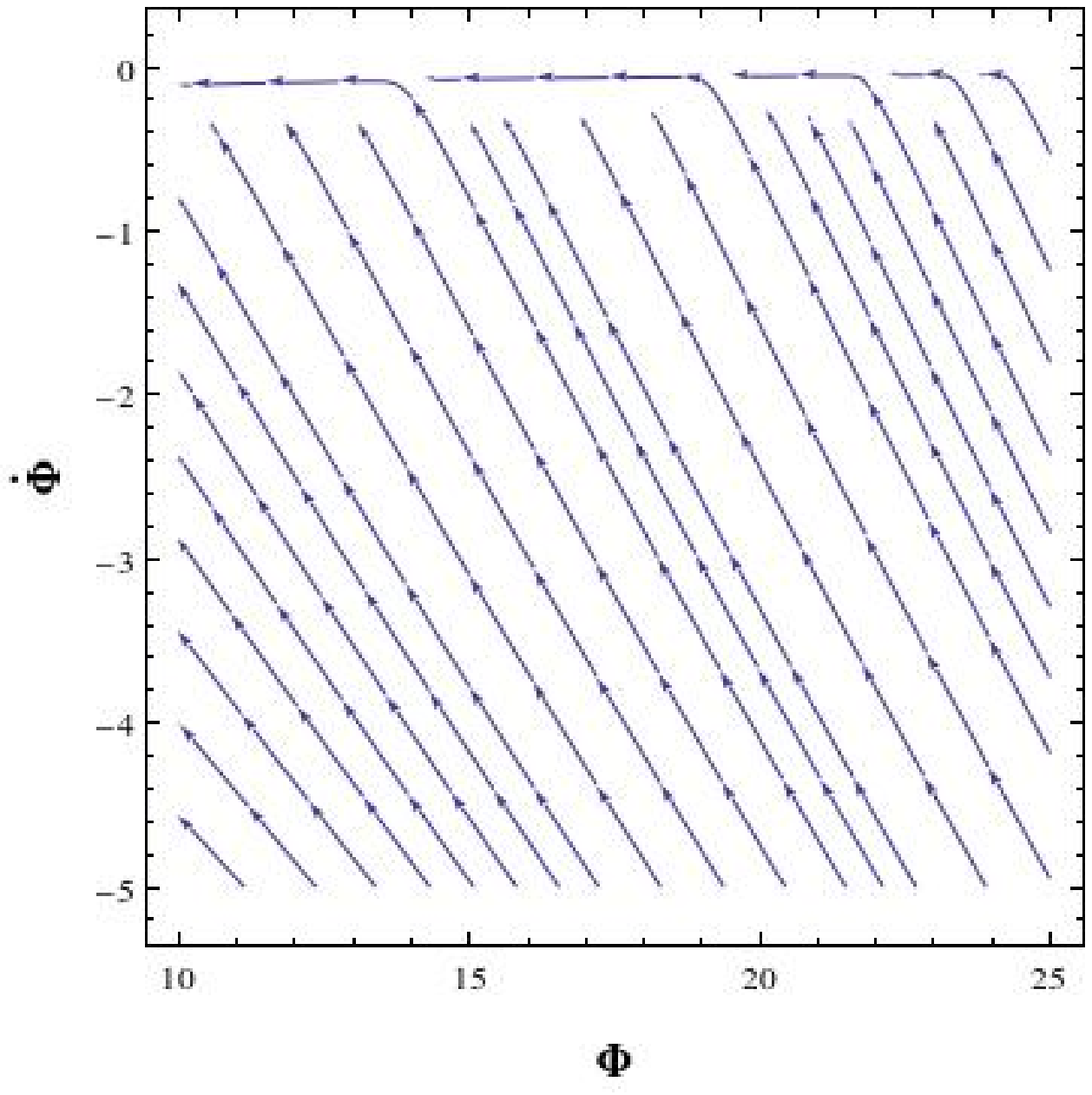}
\caption{Direction fields on the $(\Phi, \dot{\Phi})$ phase space for
positive values of $\Phi $ and various values of $\protect\beta $: (a) $%
\protect\beta =0.1$, (b) $\protect\beta =0.3$, \, (c) $\protect\beta =1$.}
\label{fig:DFPhisPositive}
\end{figure}
\end{enumerate}

We conclude that inflation happens in this scenario, but we cannot be decisive if it ends or
not. In order to achieve a definitive answer, the equations should be studied without any
approximation. This is done in the next section.


\subsection{Analysis of the complete equations ---{} Assessing the existence of an end to inflation}

In order to check if and how inflation ends in our model, we have to
analyze Eqs.~(\ref{dPhi dt}) and (\ref{dTheta dt}) with no approximations.
We have to consider a set of initial conditions for $\Phi $ and $\Theta $
that are consistent with inflation (i.e. $\Theta _{i}\ll \Phi _{i}$) and
follow the evolution of the variables. The end of inflation will be
characterized by the violation of condition $\Theta \ll \Phi $. We also
expect the dynamics of the auxiliary fields will come to an end, allowing the universe to enter a radiation-dominated era. These conditions are achieved
if we find attractors (fixed points) for Eqs.~(\ref{dPhi dt}) and (\ref%
{dTheta dt}) on which the condition $\Theta \ll \Phi $ is not satisfied.


\subsubsection{Fixed points}

The fixed points of the system are found when
\begin{align}
\frac{\Theta}{\Phi}\left(\frac{-3J+\Phi-6\beta h_{c}\Theta}{%
J+\left(\Phi-2\right)-3\beta\left(\frac{\Theta}{\Phi}\right)^{2}}\right) &
=0,  \label{FixPointPhi} \\
\frac{\Phi-J}{2\beta}-3h_{c}\Theta+2\frac{\Theta^{2}}{\Phi^{2}}\left(\frac{%
-3J+\Phi-6\beta h_{c}\Theta}{J+\left(\Phi-2\right)-3\beta\left(\frac{\Theta}{%
\Phi}\right)^{2}}\right) & =0.  \label{FixPointTheta}
\end{align}

One of the fixed points is obtained when $\Theta=0$ and $J=\Phi\Rightarrow%
\Phi=1$. However, if we start in a region where $\Phi\gg1$, then we have to
cross the critical region $\Phi=2$. This is possible if and only if $%
\Theta=0 $ at the same time when $\Phi=2$, otherwise, $J$ becomes complex.
Besides, for this very same reason, when going from $\Phi=2$ to $\Phi=1$, $%
\Theta$ has to be null. However, when $\left(\Phi,\Theta\right)=\left(2,0%
\right)$, $\dot{\Phi}=0$ and $\dot{\Theta}>0$ [see Eqs.~(\ref{dPhi dt}] and (%
\ref{dTheta dt})), showing that the trajectory tends to keep $\Phi$
unchanged and to increase the values of $\Theta$, the trajectory does
not move toward the point $\left(1,0\right)$. This is what makes the region
of the configuration space $\Phi<2$ problematic for our system, and for this
reason, this fixed point will be discarded.

We find another fixed point when
\begin{equation}
\begin{cases}
-3J+\Phi_{E}-6\beta h_{c}\Theta_{E}=0 \\
-J+\Phi_{E}-6\beta h_{c}\Theta_{E}=0%
\end{cases}%
.  \label{FixPoint}
\end{equation}
In this case,
\begin{equation}
J=0\Rightarrow\Theta_{E}=\pm\sqrt{\frac{\Phi_{E}}{2\beta}\left(\Phi_{E}-2%
\right)^{2}}  \label{ThetaEq}
\end{equation}
and
\begin{equation}
\Phi_{E}=6\beta h_{c}\Theta_{E}.  \label{PhiEq}
\end{equation}
Considering that $\Phi_{E}$ and $\beta$ have the same sign and $h_{c}$ is
positive, the minus sign on Eq.~(\ref{ThetaEq}) has to be neglected.

\begin{figure}[H]
\begin{centering}
\includegraphics[scale=0.6]{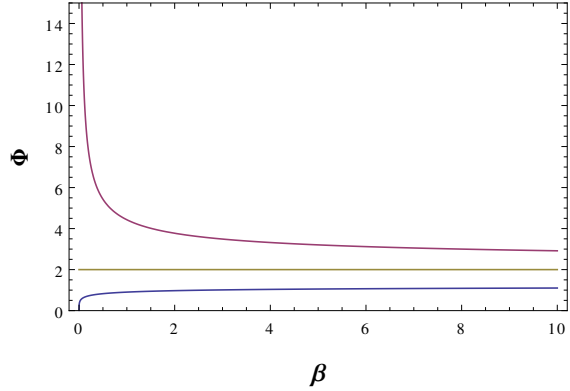}
\par\end{centering}
\caption{Real solutions $\Phi _{E}$ of Eq.~(\protect\ref{PhiEq1}) and the
function $\Phi =2$ as a function of $\protect\beta $. }
\label{fig:PhiE}
\end{figure}

Replacing Eq.~(\ref{ThetaEq}) in Eq.~(\ref{PhiEq}) shows that $\Phi _{E}$ is
found as a solution of the following equation:
\begin{equation}
\left( \Phi _{E}-\frac{6}{5}\right) \left( \Phi _{E}-2\right) ^{3}\beta =%
\frac{8\Phi _{E}^{3}}{15}.  \label{PhiEq1}
\end{equation}%
This is a fourth-order equation for $\Phi _{E}$, which presents four
solutions. Two of them are complex solutions and for this reason they will
be neglected. The other two solutions are real. One of them, however,
is bound to be less than $6/5$, for any value of $\beta $ ---{} see Fig. \ref%
{fig:PhiE}. As has been discussed before, the region $\Phi <2$ is not
suitable for our system: we are left with only one of the four solutions
---{} the upper curve in Fig. \ref{fig:PhiE}.

First, we check consistency with the Hubble function, which is rewritten
as:
\begin{equation}
h_{c}^{2}\left(\Phi_{E},\Theta_{E}\right)=\frac{1}{24}\frac{1}{\Phi_{E}^{2}}%
\left(\Phi_{E}-\frac{6}{5}\right)\left(\Phi_{E}-2\right).  \label{hcSquared}
\end{equation}
For $\Phi_{E}>2$, $h_{c}^{2}>0$ and $h_{c}$ is real and positive.

Second, we have to check if the denominator $D\left(\Phi,\Theta\right)%
\equiv J+\left(\Phi-2\right)-3\beta\left(\frac{\Theta}{\Phi}\right)^{2}$
---{} see Eqs.~(\ref{FixPointPhi}) and (\ref{FixPointTheta})
---{} does not vanish at the fixed point. We have
\begin{equation}
D\left(\Phi_{E},\Theta_{E}\right)=-\frac{\left(\Phi_{E}-2\right)\left(%
\Phi_{E}-6\right)}{2\Phi}  \label{eq:Denom}
\end{equation}
This expression shows we have a singularity if $\Phi_{E}=6$. This
singularity can be avoided if we correctly choose the value of the parameter
$\beta$. Equation (\ref{PhiEq1}) can be used to invert $\beta$ in terms of $%
\Phi_{E}$ :
\begin{equation}
\beta=\frac{8\Phi_{E}^{3}}{15\left(\Phi_{E}-\frac{6}{5}\right)\left(%
\Phi_{E}-2\right)^{3}}.  \label{eq:BetaPhiEq}
\end{equation}
If we set $\Phi_{E}=6$, then $\beta=0.375$. If we choose $\beta\neq0.375$,
then for $\Phi_{E}>2$ we have no problem of singularities with $D\left(\Phi_{E},\Theta_{E}\right)$.

In the third place, we check if the scale factor decelerates at the fixed point,
allowing inflation to cease and the universe to change from a de Sitter--type phase to a
\emph{decelerated} expansion. From Eq.~(\ref{a2dots}), we find
\begin{equation*}
\frac{\ddot{a}}{a}=-\frac{1}{48\Phi _{E}^{2}}\left( \Phi _{E}-6\right)
\left( \Phi _{E}-2\right) .
\end{equation*}%
Consequently, if $2<\Phi _{E}<6$, we have an acceleration of the scale
factor, while for $\Phi _{E}>6$ we have a deceleration. Therefore, in order
to have a suitable fixed point which could eventually lead to a good model
of inflation we are forced to choose $0<\beta <0.375$, according to Eq.~(\ref%
{eq:BetaPhiEq}).


\subsubsection{Stability of the fixed point}

At last, we have to check if the fixed point is an attractor. In principle, this
can be done by assessing the Lyapunov coefficients. These are
obtained as the eigenvalues $\lambda _{\pm }$ of the matrix
\begin{equation}
M= \left(
\begin{array}{cc}
\frac{\partial f}{\partial \Phi } & \frac{\partial f}{\partial \Theta } \\
\frac{\partial g}{\partial \Phi } & \frac{\partial g}{\partial \Theta }%
\end{array}%
\right) _{%
\begin{array}{c}
\Phi =\Phi _{E} \\
\Theta =\Theta _{E}%
\end{array}%
},  \label{M}
\end{equation}%
whose elements are calculated from Eqs.~(\ref{dPhi dt}) and (\ref{dTheta dt}).
Since we have a $2\times 2$ matrix, it is immediate that:
\begin{equation*}
\lambda _{\pm }=\frac{1}{2}\left[ \left( \frac{\partial f}{\partial \Phi }+%
\frac{\partial g}{\partial \Theta }\right) \pm \sqrt{\left( \frac{\partial f%
}{\partial \Phi }+\frac{\partial g}{\partial \Theta }\right) ^{2}-4\left(
\frac{\partial f}{\partial \Phi }\frac{\partial g}{\partial \Theta }-\frac{%
\partial g}{\partial \Phi }\frac{\partial f}{\partial \Theta }\right) }%
\right] _{%
\begin{array}{c}
\Phi =\Phi _{E} \\
\Theta =\Theta _{E}%
\end{array}%
}
\end{equation*}%
However, the partial derivatives in the above expression
diverge, making the Lyapunov coefficients analysis difficult to be
implemented. In order to circumvent this problem, we have to look for
numerical solutions and see if an attractor appears.

We performed numerical calculations in order to build the direction
fields on the $(\Phi, \dot{\Phi})$ phase space and checked the existence
of an attractor on this space. The procedure for building these curves was
the following:

\begin{itemize}
\item[--] we started with initial conditions $\left( \Phi _{i},\Theta
_{i}\right) $ and a fixed value $\beta $ and found numerical solutions $%
\left( \Phi _{i}\left( t\right) ,\Theta _{i}\left( t\right) \right) $;

\item[--] by plotting the curves $\Phi _{i}\left( t\right) $, we verified
that $\Phi _{i}\left( t\right) $ is monotonous in time, so the inverse
relation $t_{i}=t\left( \Phi _{i}\right) $ was obtained;

\item[--] from the solutions $\Theta _{i}\left( t\right) $, we found $\Theta
\left( \Phi \right) =\Theta _{i}\left( t\left( \Phi _{i}\right) \right) $;

\item[--] we derived Eq.~(\ref{dPhi dt}) with respect to time. This led
to a second order equation for $\Phi $, i.e. $\ddot{\Phi}=\ddot{\Phi}\left(
\Phi ,\dot{\Phi},\Theta ,\dot{\Theta}\right) $;

\item[--] we replaced $\ddot{\Phi}$ by $\frac{d\dot{\Phi}}{d\Phi }\dot{\Phi}$%
, $\dot{\Theta}$ by $\dot{\Theta}\left( \Phi ,\Theta \left( \Phi \right)
\right) $ and $\Theta $ by $\Theta \left( \Phi \right) $ thus obtaining $%
\frac{d\dot{\Phi}}{d\Phi }=\frac{1}{\dot{\Phi}}F\left( \Phi ,\dot{\Phi}%
\right) $, where $F\left( \Phi ,\dot{\Phi}\right) =\ddot{\Phi}\left( \Phi ,%
\dot{\Phi},\Theta \left( \Phi \right) ,\dot{\Theta}\left( \Phi ,\Theta
\left( \Phi \right) \right) \right) $;

\item[--] the direction fields on the $(\Phi,\dot{\Phi})$ phase space were plotted.
\end{itemize}

\begin{figure}[H]
\begin{centering}
\includegraphics[scale=0.6]{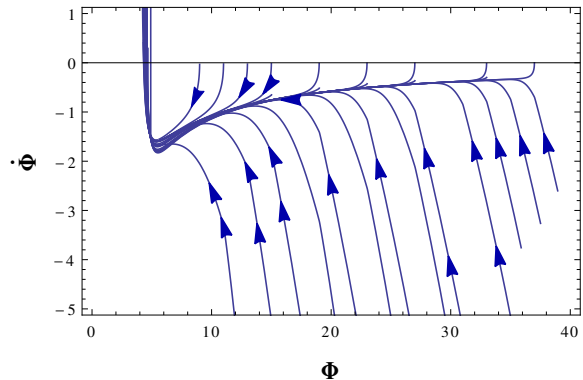}
\par\end{centering}
\caption{Direction fields on the $(\Phi,\dot{\Phi})$ phase space for
the complete set of equations with $\beta =0.1$.}
\label{fig:DFComplete}
\end{figure}

The result is presented in Fig.~\ref{fig:DFComplete} for $\beta = 0.1$.  Other values of $\beta$ produce similar graphs; for this reason, they were not displayed in the text. The graph in Fig.~\ref{fig:DFComplete} displays an attractor line along which $\Phi $ decreases
slowly (toward corresponding smaller values of $\dot{\Phi}$). So the slow-roll regime takes place even for the nonapproximate equations. As $\Phi $ approaches $\sim 5$, the
attractor line changes drastically: $\Phi $ practically stops decreasing and $\dot{\Phi}$ increases very fast. The trajectory diverges in the phase space
and no attractor can be identified. The conclusion is straightforward: there
is an inflationary period but the trajectories do not lead to an attractor;
hence, the dynamics of the auxiliary fields does not seem to end and the
model does not provide a satisfactory end for the inflationary period. No
graceful exit can be identified by the numerical solutions.

We conclude this section by stating that the Starobinsky-Podolsky model
analyzed in the Einstein frame in the Palatini formalism does not provide a
satisfactory model for inflation.


\section{Final remarks \label{sec:Final-remarks}}

This paper was dedicated to construct the scalar-multitensorial equivalent
of higher-order $f\left(R,\nabla R,\dots\right)$ theories of gravity in the
Einstein frame. The work was performed in the metric and Palatini formalisms
pointing out the differences and similarities. The main difference
between these formulations is that in the metric approach there is a clear
kinetic structure for the scalar field $\tilde{\Phi}$, whereas in the
Palatini approach this structure is absent. This is explicitly verified in
the differences between the effective energy-momentum tensors and between
the scalar field equations of both approaches. In addition, we also
thoroughly studied the particular case of the Starobinsky-Podolsky model in the 
Einstein frame and characterized its effective energy-momentum tensor in
terms of a fluid description, where a shearless imperfect fluid is obtained
in the metric approach, while a perfect fluid is obtained in Palatini
formalism. This completes the development started in Ref. \cite{PRD2016}.

An important point to be emphasized is that although the higher-order 
$f\left(R,\nabla R,\dots\right)$ theories have an apparently similar form to
the $f\left(R\right)$ theories in the Einstein frame, they differ substantially
due to the structure of the potential $\tilde{U}$. While in $f\left(R\right)$
theories the potential $\tilde{U}$ depends only on the scalar field, in
higher-order models, $\tilde{U}$ has a much more complex structure depending
on the extra fields $\left\{
\Phi,\phi_{\mu},\dots,\phi_{\mu_{1}\dots\mu_{n}}\right\} $ and its covariant
derivatives. However, even taking into account the complexity of the
potential $\tilde{U}$, the $f\left(R,\nabla R,\dots\right)$-like theories are
simplified considerably when rewritten in Einstein frame. This is
particularly true in the situation where the higher-order terms are small
corrections to the Einstein-Hilbert (or Starobinsky) action and in this case
the potential $\tilde{U}$ can be treated in a perturbative way.

There are several cases where it is convenient to describe the $f\left(
R,\nabla R,\dots\right) $ theories in Einstein frame. In particular, an
important case to be analyzed is the study and description of ghosts.
Following an approach analogous to that in Ref. \cite{Hindawi1996}, one can
study in which higher-order theories and under which conditions the
pathologies involving ghosts are avoided.

Another interesting case takes place in inflationary cosmology. Usually, inflationary models based on
modified gravity are more easily described in the Einstein frame. For example,
Starobinsky's inflation \cite{Starobinsky1980} is widely studied through its
scalar-tensorial version. In Sec.~\ref{sec:Inflation}, we
explored the possibility of generating inflation through
Starobinsky-Podolsky gravity in the Palatini formalism. We showed that the
extra (vector) field is able to engender an inflationary regime for a wide
range of initial conditions. However, this regime does not end in a satisfactory
way: it does not lead to a hot-big-bang type of dynamics. Some possibilities to
circumvent this problem are to include even higher-order terms, to introduce extra
new fields or even to consider nonminimal couplings. Finally, it is worth mentioning
that a similar but more involved analysis was done in \cite{StaPodInf}; this paper considered Starobinsky-Podolsky inflation in the \emph{metric} formalism and showed a graceful exit is possible for the model.

\acknowledgements

R.R.C. is grateful for the hospitality of Robert H. Brandenberger and the people
at McGill Physics Department (Montreal, Quebec, Canada) where this work was
initiated. This study was financed in part by Coordena\c c\~ao de Aperfei\c coamento de 
Pessoal de N\'ivel Superior ---{} Brasil (CAPES) ---{} Finance Code 001. L.G.M. is 
grateful to CNPq-Brazil for partial financial support.


\end{document}